%% file: arxiv_GLA.tex
\documentclass[]{article}

\usepackage{amssymb}
\usepackage{amsmath}
\usepackage{graphicx}
\usepackage{epsfig}
\usepackage{float}
\usepackage{subfigure}
\usepackage{psfrag}
\usepackage[utf8x]{inputenc}
\usepackage{color}
\usepackage{array}
\usepackage{dsfont}
\usepackage{algorithm,algorithmic}
\usepackage{tikz}
\usepackage{listings}
\usepackage{epsfig}
\usetikzlibrary{arrows,decorations.pathmorphing,backgrounds,fit,positioning,shapes.symbols,chains,mindmap,trees,snakes,calc}
\usepackage{multirow}
\usepackage{booktabs}
\usepackage{a4wide}
\usepackage{cleveref}

\definecolor{bleuONERA}{RGB}{16,97,169}
\definecolor{grisONERA}{RGB}{64,64,66}

\newtheorem{remark}{Remark}

\providecommand{\mimo}[0]{\textbf{MIMO}~}

\providecommand{\lti}[0]{\textbf{LTI}~}

\providecommand{\irka}[0]{\textbf{IRKA}~}

\providecommand{\flistia}[0]{\textbf{FL-ISTIA}~}

\providecommand{\darpo}[0]{\textbf{DARPO}~}

\providecommand{\ie}[0]{\emph{i.e.}~}
\providecommand{\etc}[0]{\emph{etc.}~}
\providecommand{\eg}[0]{\emph{e.g.}~}

\newenvironment{eq}{\everymath {\displaystyle \everymath{ }} \equation}{ \endequation} %
\DeclareMathOperator*{\rank}{\mathbf{rank}}

\providecommand{\abs}[1]{\left\lvert #1 \right\rvert} %
\providecommand{\norml}[1]{\left\lVert #1 \right\rVert} %
\providecommand{\norm}[1]{|| #1 ||} %
\providecommand{\eval}[2]{\left.#1\right\rvert_{#2}} %
\providecommand{\pare}[1]{\left(#1\right) } %
\providecommand{\x}[0]{\mathbf{x}} %
\providecommand{\dx}[0]{\mathbf{\dot{x}}} %
\providecommand{\xr}[0]{\mathbf{\hat{x}}} %
\renewcommand{\u}{\mathbf{u}} %
\providecommand{\y}[0]{\mathbf{y}} %
\providecommand{\yr}[0]{\mathbf{\hat{y}}} %
\providecommand{\lv}[0]{\mathbf{l}} %
\providecommand{\rv}[0]{\mathbf{r}} %
\providecommand{\wv}[0]{\mathbf{w}} %
\providecommand{\vv}[0]{\mathbf{v}} %
\providecommand{\cv}[0]{\mathbf{c}} %
\providecommand{\bv}[0]{\mathbf{b}} %

\providecommand{\Hrealr}[0]{\mathcal{\hat{S}}} %
\providecommand{\Htranr}[0]{\mathbf{\hat{H}}} %
\providecommand{\Er}[0]{{\hat{E}}} %
\providecommand{\Ar}[0]{{\hat{A}}} %
\providecommand{\Br}[0]{{\hat{B}}} %
\providecommand{\Cr}[0]{{\hat{C}}} %
\providecommand{\Hreal}[0]{\mathcal{S}} %
\providecommand{\Htran}[0]{\mathbf{H}} %
\providecommand{\Gtran}[0]{\mathbf{G}} %
\providecommand{\E}[0]{{E}} %
\providecommand{\A}[0]{{A}} %
\providecommand{\B}[0]{{B}} %
\providecommand{\C}[0]{{C}} %
\providecommand{\D}[0]{{D}} %

\providecommand{\LL}[0]{{\mathds L}} %
\providecommand{\sLL}[0]{{\mathds L_\sigma}} %

\providecommand{\Htwo}[0]{{\mathcal{H}_{2}}} %
\providecommand{\Htwow}[0]{{\mathcal{H}_{2,\Omega}}} %
\providecommand{\Hinf}[0]{{\mathcal{H}_{\infty}}} %
\providecommand{\Linf}[0]{{\mathcal{L}_{\infty}}} %

\providecommand{\Cplx}[0]{\mathbb{C}} %
\providecommand{\Real}[0]{\mathbb{R}} %
\providecommand{\vectortwo}[2]{ \left[\begin{array}{c} #1 \\ #2 \end{array}\right] } %
\newcommand{\udisk}{\mathcal{D}}
\newcommand{\ucir}{\partial \mathcal{D}}
\newcommand{\udiskcomp}{\overline{\mathcal{D}}}
\newcommand{\cp}{\mathbb{C}_+}

\newcommand{\chinf}{\mathcal{H}_\infty (\cp)}
\newcommand{\dhinf}{\mathcal{H}_\infty (\udiskcomp)}

\begin{document}


\title{Interpolatory Methods for Generic BizJet Gust Load Alleviation Function}

\author{C. Poussot-Vassal, P. Vuillemin, O. Cantinaud and F. S\`eve
\thanks{C. Poussot-Vassal and P. Vuillemin are with Universit\'e de Toulouse, F-31055 Toulouse, France. Contact:  \texttt{charles.poussot-vassal@onera.fr}. O. Cantinaud and F. S\`eve are with Dassault-Aviation, France.}
\thanks{This work has been funded within the frame of the Joint Technology Initiative JTI Clean Sky 2, AIRFRAME Integrated Technology Demonstrator platform AIRFRAME ITD (contract N. CSJU-CS2-GAM-AIR-2014-15-01 Annex 1, Issue B04, October 2nd, 2015) being part of the Horizon 2020 research and Innovation framework program of the European Commission.}
}
\maketitle


\begin{abstract}
The paper's main contribution concerns the use of interpolatory methods to solve end to end industrial control problems involving complex linear dynamical systems. More in details, contributions show how the rational data and function interpolation framework is a pivotal tool \emph{(i)} to construct (frequency-limited) reduced order  dynamical models appropriate for model-based control design and \emph{(ii)} to accurately discretise controllers in view of on-board computer-limited implementation. These contributions are illustrated along the paper through the design of an active feedback gust load alleviation function, applied on an industrial generic business jet aircraft use-case. The closed-loop validation and performances evaluation are assessed through the use of an industrial dedicated simulator and considering certification objectives. Although application is centred on aircraft applications, the method is not restrictive and can be extended to any linear dynamical systems.
\end{abstract}

\section{Introduction}
\label{sec:intro}
\input{sec-intro}

\section{Gust load oriented modelling}
\label{sec:modeling}
\input{sec-modeling}

\section{Preliminaries in rational interpolation and approximation}
\label{sec:interp}

\input{sec-interp}

\section{Interpolatory-driven aero-servoelastic aircraft gust load oriented reduced modelling}
\label{sec:model}
\input{sec-model}

\section{Interpolatory-driven sampled-time feedback gust load alleviation function computation}
\label{sec:control}
\input{sec-control}

\section{Conclusions}
\label{sec:conclusions}
\input{sec-conclusions}


\end{document}

%% file: sec-intro.tex
\subsection{General aircraft context}

Aircraft mobility plays an important role in our life style and societal organisation. As this transportation mean is facing severe environmental and societal challenges (\eg global warming and CO$_2$ emissions), it is the role of researchers to provide innovative answers and industry-oriented tools to address them. These solutions should   fulfil the safety and design requirements and be numerically efficient and simple to implement within the aircraft industrial design value chain. Among others scientific disciplines, it is well admitted that civil aircraft industry relies on dynamical systems theory, linear algebra and computational sciences to address these issues. In light of these statements, recent developments in these scientific communities may have a major impact in the overall aircraft conception and exploitation enhancement \cite{Amsallem:2011,Willcox:2002}. Hopefully, it can also be a substantial lever in the \emph{civil aviation footprint reduction}. This is one of the main expectation of this work. 

In order to reach these ambitious objectives, the paper highlights the pivotal role and relevance of \emph{interpolatory methods} \cite{AntoulasBook:2020,AntoulasSurvey:2016} in the context of \emph{linear time invariant (\textbf{LTI}) large-scale systems} \cite{SaadBook:2000}. 

\subsection{Generic business jet aircraft context}

More specifically, through a complete industrial gust load alleviation control design and validation problem applied to a generic business jet aircraft, we show how rational function interpolation is an appealing ingredient for engineers. It is used \emph{(i)} to construct \emph{reduced order dynamical models} appropriate for feedback control design and \emph{(ii)} to compute more accurate \emph{sampled-time} gust load dynamical controller functions that can be implemented in a constrained on-board computer. The complete closed-loop stability and performance analysis is done through a dedicated \emph{industrial simulator} to assess the approach. 

\subsection{Generic business gust load alleviation feedback control objective}

An important aircraft design criteria concerns the \emph{gust load envelope} that should not reach some limit to ensure structure integrity. To this aim, it is standard considering vertical gusts wavelengths, modelled through the following "1-cosine" profile
\begin{equation}
\mathbf w_g(t) = 
\left\{
\begin{array}{ll}
\dfrac{W}{2}\pare{1-\cos\pare{\dfrac{\pi V}{L}t}} & \text{ for $0\leq t \leq \dfrac{2L}{V}$}\\
0 & \text{ for $0>\dfrac{2L}{V}$}
\end{array}
\right. ,
\label{eq:gust}
\end{equation}
where $W$ is the gust velocity (in feet), $L$ is the gust wavelength (in meters) and $V$ is the aircraft true airspeed (in meter per seconds)\footnote{Typical values for these parameters are provided by authorities and results in hundreds of different gust configurations.}. The \emph{gust load envelope} is simply the worst case load responses along the wing span in reaction to the set of time-domain vertical wind gust profiles \cref{eq:gust} affecting the aircraft. Similarly to aerodynamical effects inducing vibrations \cite{MeyerIFASD:2017}, these gusts represent typical phenomena the aircraft might face during its exploitation. Prior any flight test, aircraft manufacturers should guarantee authorities that they can be handled by the system.

In the preliminary conception step (before considering any control functions), the aircraft is designed by experts so that the wings should support a given nominal load envelope, dictated by physical considerations such as desired aircraft manoeuvrability, gust, and many other manufacturing constraints. The larger the supported loads are, the larger the structural stiffeners and mass reinforcements should be. The aircraft mass is consequently bigger and its consumption during flight increased. In this context, the \emph{Gust Load Alleviation} (GLA) control function plays an important role in the aircraft conception: it is aimed at lowering the loads envelope and thus at reducing the aircraft overall mass, consumption and emissions \cite{Alama:2015,Gadient:2012}. In this work, the gain brought by the GLA, on the \emph{worst upward gain} is denoted by $\mathcal E(x_i)$, is computed as 
\begin{equation}
\mathcal E(x_i) = \max_{
\mathbf w_g \in \cal W\\
} \norml{\dfrac{\mathbf z_{\text{loads}}(x_i)-\mathbf z^{\text{(GLA)}}_{\text{loads}}(x_i)}{\mathbf z_{\text{loads}}(x_i)}}_\infty,
\label{eq:envelope}
\end{equation}
where $x_i$ (in meters) is the wing location points where the loads are computed (here 
five locations $x_i=[x_1,x_2,x_3,x_4,x_5]$m) and where $\cal W$ is the set of gust profiles as in \eqref{eq:gust}. Then, $\mathbf z_{\text{loads}}(x_i)$ denotes the load at location $x_i$ for the baseline aircraft and $\mathbf z^{(GLA)}_{\text{loads}}(x_i)$ denotes the one when the \emph{Gust Load Alleviation} (GLA) control function is activated. Both responses are  obtained when the aircraft model is fed by \cref{eq:gust} (see also \cref{sec:modeling}).

\subsection{Generic business gust load alleviation feedback industrial constraints}

In an industrial value chain, the GLA function is traditionally designed after the flight controller (focusing on handling qualities). Then, one requirement is that GLA control should not impact the nominal (low frequencies) flight performances and focus on the gust phenomena and load envelope only. In addition, as the control functions should be implemented in a limited sampled-time computer, with a constrained material architecture (sensors and actuators limitations, sampling limitations, delays in the loop, \etc), the discretisation step should also be accurately taken into account before any implementation and performance evaluation. 


\subsection{Main results and paper organisation}

Given the above considerations and objectives, the rest of the paper is focused on the central role of the \emph{interpolation} applied to the \emph{gust load alleviation control} problem. The paper is organised as follows. The gust load-oriented aircraft modelling is detailed in \cref{sec:modeling}, leading to a set of medium-scale irrational models. As the main mathematical tool involved here, reminders on the interpolation framework are recalled in \cref{sec:interp}. Its application for the construction of rational reduced order models set, suitable for linear control design, is described in \cref{sec:model}. Then, after briefly describing the continuous-time GLA controller synthesis and the continuous and sampled-time interconnection framework, the controller \emph{discretisation} through the interpolation framework is detailed in \cref{sec:control}. This last point stands as a methodological contribution of the work. Finally, conclusions are given in \cref{sec:conclusions}, illustrating first the efficiency of the proposed process to alleviate gust-driven loads in an industrial application context, and second, to summarise the main contributions that can be applied to any kind of linear dynamical models.

Although the paper is centred around the aircraft gust load function, the proposed approach is readily applicable to any other aircraft control problems (including aircraft flight quality, vibrations, \etc functions). It is also valid to (m)any linear time invariant dynamical use-cases. Indeed, the presented \emph{interpolatory framework} versatility allows to address a large number of engineering applications. 

\subsection{Notations}

Let us denote by $\Real$ the set of real numbers, $\Cplx$ the set of complex numbers, $\cp$ ($\Cplx_-$) the open right (left) half plane, $\udisk$ the open unit disk, $\ucir$ its boundary and $\udiskcomp$ the complementary of the closed unit disk, respectively. The complex variable is given by $\imath=\sqrt{-1}$. Let $\mathcal{L}_2(\mathcal{I})$ ($\mathcal{I} = \imath\mathbb{R}$ or $\ucir$) be the set of functions that are square integrable on $\mathcal{I}$. Let $\mathcal{H}_2(\mathcal{\udisk})$ (resp. $\mathcal{H}_2(\mathcal{\udiskcomp})$) be the subset of $\mathcal{L}_2(\ucir)$ containing the functions analytic in $\udisk$ (resp. $\udiskcomp$). Let $\Htwo(\cp)$, shortly $\Htwo$,  (resp. $\mathcal{H}_2(\Cplx_-)$) be the subset of $\mathcal{L}_2(\imath\Real)$ containing the functions analytic in $\cp$ (resp. $\Cplx_-$). Similarly, let $\mathcal{L}_\infty (\mathcal{I})$ ($\mathcal{I} = \imath\mathbb{R}$ or $\ucir$) be the set of functions that are bounded on $\mathcal{I}$. Let $\mathcal{H}_\infty(\mathcal{\udisk})$ (resp. $\mathcal{H}_\infty(\mathcal{\udiskcomp})$) be the subset of $\mathcal{L}_\infty(\ucir)$ containing the functions analytic in $\udisk$ (resp. $\udiskcomp$) and $\mathcal{H}_\infty(\cp)$, shortly $\Hinf$, the subset of $\mathcal{L}_\infty(\imath\Real)$ of functions analytic in $\cp$. 
The Fourier transform of a time-domain signal $v \in \mathcal{L}_2(\mathbb{R})$ is denoted by $\overline{v} = \mathcal{F}(v)$.

%% file: sec-modeling.tex
\begin{figure}[htbp]
  \centering
  \scalebox{.8}{\input{figures/systemModelLoad_delay}}
  \caption{Closed-loop architecture of the GLA problem. The complete aeroservoelastic dynamical aircraft models $\{\Gtran_i\}_{i=1}^{n_s}$ include the "Flight controller", "Actuators", "Sensors" and "Computational delay $\tau_m$". The "GLA controller" is GLA function to be computed. Signals $\mathbf w$, $\u$, $\mathbf z$ and $\y$ denote the exogenous inputs, control inputs, performance outputs and measurements, respectively. Then $h$ denotes the sampling time usd in the GLA function.}
  \label{fig:global}
\end{figure}
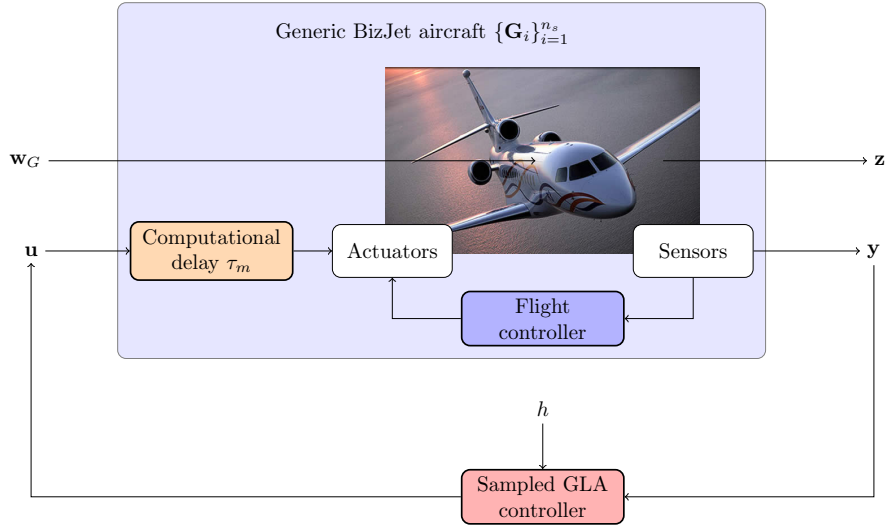

\cref{fig:global} illustrates the considered feedback control loop architecture, where the  sampled-time GLA function denoted as "GLA controller" is to be computed and validated. One may note that the actuators, sensors, flight controller and computational delay are considered as given and included in the considered generic business jet (BizJet) aircraft model set, referred to as $\{\Gtran_i\}_{i=1}^{n_s}$, and detailed hereafter. 

In \cref{sub-earoservo}, the large-scale aeroservoelastic models $\{\Gtran_i\}_{i=1}^{n_s}$, as provided by physics and loads teams, is described. Its input-output simplification is then done in \cref{sub-glaMdl}, leading to the the set of irrational models $\{\Htran_i\}_{i=1}^{n_s}$.

\subsection{Aeroservoelastic aircraft modelling}
\label{sub-earoservo}

To guarantee the safety and performance levels, GLA function is developed using very accurate environment and models, issued from aeroservoelastic modelling teams \cite{Push:2019,QueroAerospace:2019,QueroPhD:2017,Kier:2009}.

Following \cref{fig:global}, $\{\Gtran_i\}_{i=1}^{n_s}$ denotes the gust load oriented dynamical models family complex-valued functions. Each $\Gtran_i$ ($i=1,\dots,n_s$) represents a \emph{linear time invariant dynamical model} of the aircraft, evaluated at a given flight and mass condition. Its construction is a know-how of the aircraft manufacturer and usually results from multiple steps performed by different experts from fluid mechanics, structural, tests \etc teams. Each model $\{\Gtran_i\}_{i=1}^{n_s}$ is described by the following realisation denoted by $\{\mathcal G_i\}_{i=1}^{n_s}$
\begin{equation}
\{\mathcal G_i\}_{i=1}^{n_s} : \left\{
\begin{array}{rcl}
\dx^{(i)}(t) &=& \A^{(i)}\x^{(i)}(t) + \B^{(i)}\vectortwo{\mathbf w_G(t)}{\u(t)}\\
\vectortwo{\y(t)}{\mathbf z(t)} &=& \C^{(i)}\x^{(i)}(t) + \C_d^{(i)}\x^{(i)}(t-\tau_m)
\end{array}
\right.
\label{eq:model1ssG}
\end{equation}
where $\x^{(i)}(t)\in\Real^{M_i}$, $\u(t)\in\Real^{n_u}$, $\mathbf w_G(t)\in\Real^{n_{wG}}$, $\y(t)\in\Real^{n_y}$ and $\mathbf z(t)\in\Real^{n_z}$ are the internal variables, input, exogenous input, output and performance output signals, respectively. The dimensions $n_u$, $n_{wG}$, $n_y$ and $n_z$ are constant for all $n_s$ models, while $M_i$ depends on the model configuration ($i=1,\dots,n_s$). Finally, $\tau_m\in\Real_+$ represents the computational delay needed to capture the $\y$ signal (and to compute the - to be designed - GLA control law). 

In the considered use-case, $\u$ gathers three control inputs, the horizontal tail, the inner and outer ailerons ($n_u=3$). Exogenous input $\mathbf w_G$ gathers $\mathbf \delta_{mc}^{\star}$, the aircraft horizontal tail deflection given by the pilot, the gust disturbance and its first and second derivatives, applied at three different locations of the aircraft fuselage ($n_w=7$). This vector reads
\begin{equation}
\begin{array}{rcl}
\mathbf w_G(t) &=& 
\bigg[
\mathbf \delta_{mc}^{\star}(t), 
\underbrace{\mathbf w_g(t), \mathbf{\dot w}_g(t), \mathbf{\ddot w}_g(t)}_{\text{aircraft front}}, 
\underbrace{\mathbf w_g(t-\tau_1^{(i)}), \mathbf{\dot w}_g(t-\tau_1^{(i)}), \mathbf{\ddot w}_g(t-\tau_1^{(i)})}_{\text{aircraft middle}}, \dots \\
& & \underbrace{\mathbf w_g(t-\tau_2^{(i)}), \mathbf{\dot w}_g(t-\tau_2^{(i)}), \mathbf{\ddot w}_g(t-\tau_2^{(i)})}_{\text{aircraft rear}}
\bigg]^T,
\end{array}
\label{eq:w}
\end{equation}
where the same gust (position, velocity and acceleration) enters in the model at the front, the middle and the rear, with a delay $\tau_1^{(i)}\in\Real_+$ and $\tau_2^{(i)}\in\Real_+$, depending on the aircraft velocity ($i=1,\dots ,n_s$). The only considered measurement $\y$ is the aircraft angle of attack at the nose level ($n_y=1$). Finally, the performances $\mathbf z$ gathers outputs being the tracking signal error between the load factor without ($\mathbf n_{z}^{\star}$) and with ($\mathbf n_{z}$)  GLA controller $\mathbf n_{z}^{\star}-\mathbf n_{z}$ and the loads envelope $\mathbf z_{\text{loads}}(x_i)$ ($n_z=6$). The $\A^{(i)}$, $\B^{(i)}$, $\C^{(i)}$ and $\C_d^{(i)}$ matrices are real and of appropriate dimensions.

\subsection{Toward gust load control-oriented models}
\label{sub-glaMdl}

As signals in \cref{eq:w} are repeated and linked together, it can problematic to deal with in a control design setup. One solution is to merge them in order to deal with $\mathbf w(t) = \left[\mathbf \delta_{mc}^{\star}(t), \mathbf w_G(t)\right]^T$ instead of \cref{eq:w}. This can be obtained by applying the following simple transformation
\begin{equation}
\left[
\begin{array}{l}
\mathbf \delta_{mc}^{\star}(t)\\
\mathbf w_g(t)\\ \mathbf{\dot w}_g(t)\\ \mathbf{\ddot w}_g(t)\\ 
\mathbf w_g(t-\tau_1^{(i)})\\ \mathbf{\dot w}_g(t-\tau_1^{(i)})\\ \mathbf{\ddot w}_g(t-\tau_1^{(i)})\\
\mathbf w_g(t-\tau_2^{(i)})\\ \mathbf{\dot w}_g(t-\tau_2^{(i)})\\ \mathbf{\ddot w}_g(t-\tau_2^{(i)})\\
\end{array}
\right] = 
\left[ 
\begin{array}{cl}
1 & 0\\
0 & 1\\
0 & s \\
0 & s^2 \\
0 & e^{-\tau_1^{(i)}s}\\
0 & se^{-\tau_1^{(i)}s} \\
0 & s^2e^{-\tau_1^{(i)}s} \\
0 & e^{-\tau_2^{(i)}s}\\
0 & se^{-\tau_2^{(i)}s} \\
0 & s^2e^{-\tau_2^{(i)}s} \\
\end{array}
\right]
\vectortwo{\mathbf \delta_{mc}^{\star}(t)}{\mathbf w_G(t)},
\label{eq:wtransform}
\end{equation}
where $s$ denotes the Laplace variable. By merging \cref{eq:model1ssG} and \cref{eq:wtransform}, one can now re-construct
each model in a realisation form $\{\Hreal_i\}_{i=1}^{n_s}$ as
\begin{equation}
\{\Hreal_i\}_{i=1}^{n_s} : \left\{
\begin{array}{rcl}
\E^{(i)}\dx^{(i)}(t) &=& \A^{(i)}_0\x^{(i)}(t)+\A_1^{(i)}\x^{(i)}(t-\tau_1^{(i)})+\A_2^{(i)}\x^{(i)}(t-\tau_2^{(i)}) \\ && + \B_g^{(i)}\vectortwo{\mathbf w(t)}{\u(t)}\\
\vectortwo{\y(t)}{\mathbf z(t)} &=& \C^{(i)}_0\x^{(i)}(t) +\C^{(i)}_1\x^{(i)}(t-\tau_m) 
\end{array}
\right.
\label{eq:model1ss}
\end{equation}
where $\x^{(i)}(t)\in\Real^{N_i}$, $\u(t)\in\Real^{n_u}$, $\mathbf w(t)\in\Real^{n_w}$, $\y(t)\in\Real^{n_y}$ and $\mathbf z(t)\in\Real^{n_z}$ are the internal variables, input, exogenous input, output and performance output signals, respectively. With this new form $n_u$, $n_y$ and $n_z$ are unaffected, while $n_w=2$ (instead of 10 in \cref{eq:w}) and $N_i=M_i+6$, due to the double derivative and delay structure added. The $\E^{(i)}$, $\A_0^{(i)}$, $\A_1^{(i)}$, $\A_2^{(i)}$, $\B_g^{(i)}$, $\C_0^{(i)}$ and $\C_1^{(i)}$ matrices are real and of appropriate dimensions. 

\begin{remark}[About the internal delays $\tau_1^{(i)}$ and $\tau_2^{(i)}$, and the $\E^{(i)}$ matrix rank] \label{rmq:delayGust}
In the considered use-case, the fuselage is subdivided in three patches. Consequently, the gust vertical displacement signal $\mathbf w_g(t)$ given in \cref{eq:gust}, enters the model at $t$, $t+\tau_1^{(i)}$ and $t+\tau_2^{(i)}$ ($0<\tau_1^{(i)}<\tau_2^{(i)}$). To consider one single gust input signal (instead of three delayed), internal delay are added in the model. Similarly, to be able to accurately compute the loads along the wings, the model must takes as gust input, its vertical displacement as in \cref{eq:gust}, and its first and second derivatives as in \cref{eq:w} \cite{QueroAerospace:2019}. Once again, to limit the model number of input disturbances to one instead of three (position, velocity and acceleration), the first and second derivatives of \cref{eq:gust} are embedded in the model, leading to a descriptor form as in \cref{eq:model1ss}. More specifically, in our case, the $\E^{(i)}$ matrix  rank is then equal to $N_i-6$: two rank loss per fuselage patch (one for the first and one for the second derivative).
\end{remark}

Following $\{\Hreal_i\}_{i=1}^{n_s}$ given in equation \cref{eq:model1ss}, the gust load oriented model is a family of dynamical models which transfer function $\{\Htran_i\}_{i=1}^{n_s}$, from $\u$ and $\mathbf w$ to $\y$ and $\mathbf z$, reads
\begin{eq}
\Htran_i(s) = \pare{\C^{(i)}_0 + \C^{(i)}_1e^{-\tau^{(i)}_m s}} \pare{s\E^{(i)} - \A^{(i)}_0-\A^{(i)}_1e^{\tau^{(i)}_1s}-\A^{(i)}_2e^{\tau^{(i)}_2s} }^{-1}\B^{(i)}.
\label{eq:model1tf}
\end{eq}

%% file: figures/systemModelLoad_delay.tex
\pgfdeclarelayer{background}
\pgfdeclarelayer{foreground}
\pgfsetlayers{background,main,foreground}

\tikzstyle{actProp}  = [draw=black, fill=white,text width=5em,text centered,minimum height=2.5em,rounded corners]
\tikzstyle{contPropFLIGHT} = [draw=black,thick, fill=blue!30,text width=7em,text centered,minimum height=2.5em,rounded corners]
\tikzstyle{contPropLOAD} = [draw=black,thick, fill=red!30,text width=7em,text centered,minimum height=2.5em,rounded corners]
\tikzstyle{delayProp} = [draw=black,thick, fill=orange!30,text width=7em,text centered,minimum height=2.5em,rounded corners]
\tikzstyle{ann}      = [above, text width=5em]
\def\blockdist{2.5cm}

\begin{tikzpicture}
    \node (sys) {};
    \path (sys) node (image) []  {{\includegraphics[width=.35\columnwidth]{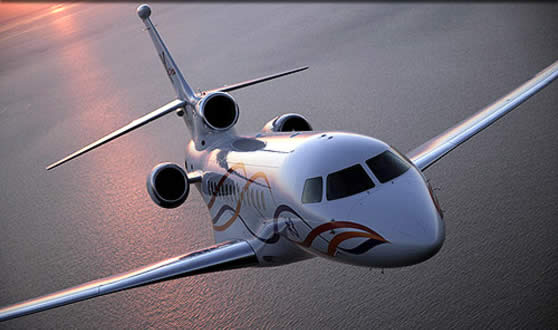}}};
    \path (sys)+(-\blockdist,-1.5cm) node (actuator) [actProp] {Actuators};
    \path (sys)+(\blockdist,-1.5cm) node (sensor) [actProp] {Sensors};
    \path (actuator.west)+(-2cm,0)  node (controllerDelay) [delayProp] {Computational delay $\tau_m$};
   
    \path (sensor.0)+(2cm,0) node (outputY) [] {$\y$};
    \draw [->] (sensor.0) -- (outputY);
    \path (sys)+(5.6cm,0) node (outputZ) [] {$\mathbf z$};
    \draw [->] (sys)+(2cm,0) -- (outputZ);
    \path (controllerDelay)+(-3cm,0cm)  node (inputU) [] {$\u$}; 
    \draw [->] (inputU) -- (controllerDelay.west);
    \draw [->] (controllerDelay.east) -- (actuator.west);
    \path (sys)+(-8.6cm,0)  node (inputW) [] {$\mathbf w_G$};
    \draw [->] (inputW) -- (sys);

    \path (sys.south)+(0,-2.5cm)  node (controllerFlight) [contPropFLIGHT] {Flight \\ controller};
    \draw [->] (sensor.-90) |- (controllerFlight.east);
    \draw [->] (controllerFlight.west) -| (actuator.-90);

    \path (controllerFlight.south)+(0cm,-2.5cm)  node (controllerLoad) [contPropLOAD] {Sampled GLA \\ controller};
    \draw [->] (outputY.-90) |- (controllerLoad.east);
    \draw [->] (controllerLoad.west) -| (inputU.-90);
    \path (controllerLoad.north)+(0cm,1cm)  node (h) [] {$h$};
    \draw [->] (h.south) -| (controllerLoad.north);
    
    \path (sys.north)+(-2cm,2) node (globalModel) {Generic BizJet aircraft $\{\Gtran_i\}_{i=1}^{n_s}$};
    \begin{pgfonlayer}{background}
        \path (controllerDelay.west |- globalModel.north)+(-0.2,0.2) node (a) {};
        \path (controllerFlight.south -| sensor.east)+(0.2,-0.2) node (b) {};
        \path [fill=blue!10,rounded corners, draw=black!50] (a) rectangle (b);
    \end{pgfonlayer}
\end{tikzpicture}

%% file: sec-interp.tex
The entire proposal relies on a specific use of the model interpolation tools: namely, the \emph{Loewner} and the \emph{optimal (frequency-limited) $\Htwo$ frameworks}. The Loewner framework is recalled in its general form, in \cref{ssec:loe}. The $\Htwo$  model order dimension reduction methods is recalled in \cref{ssec:modred}, \cref{ssec:h2} and \cref{ssec:h2w}. In the following subsections, indices and exponents in \cref{eq:model1ss} and \cref{eq:model1tf} are dropped for simplicity.

\subsection{Reminder of the Loewner framework}
\label{ssec:loe}

The main elements of the Loewner framework are recalled thereafter in the multi-input multi-output (\textbf{MIMO}) general square case. For a complete description readers may refer to \cite{AntoulasSurvey:2016,Mayo:2007}, and to \cite{Antoulas:2016} for insight in the rectangular case. Under mild considerations, the Loewner approach is a data-driven method aimed at building a rational descriptor \lti dynamical model $\Htran^m$ of dimension $m$  which interpolates given complex data, here generated by a model $\Htran$. Let us be given  the left or row data, and the right or column data:
\begin{eq}
\left.
\begin{array}{c}
(\mu_j,\lv_j^H,\vv_j^H) \\
\text{for $j=1,\dots,m$}
\end{array}
\right\}
\text{~~and~~}
\left\{
\begin{array}{c}
(\lambda_i,\rv_i,\wv_i) \\
\text{for $i=1,\dots ,m$}
\end{array}
\right. ,
\label{eq:loewnerInput}
\end{eq}
where $\vv_j^H=\lv_j^H\Htran(\mu_j)$ and $\wv_i=\Htran(\lambda_i)\rv_i$, with $\lv_j\in\Cplx^{n_y\times 1}$, $\rv_i\in\Cplx^{n_u\times 1}$, $\vv_j\in\Cplx^{n_u\times 1}$ and $\wv_i\in\Cplx^{n_y\times 1}$. The $\wv_i$ and $\vv_j$ are the complex measurement data at the points $\mu_j$ and $\lambda_i$, along the tangential directions $\lv_j$ and $\rv_i$. In addition, let us assume that $\lambda_i\in\Cplx$ and $\mu_j\in\Cplx$ represent a set of distinct interpolation points  $\{ z_k\}_{k=1}^{2m} \subset \Cplx$ which is split in two equal subsets as
\begin{equation}
\{z_k\}_{k=1}^{2m} =\{\mu_j\}_{j=1}^{m} \cup \{\lambda_i\}_{i=1}^{m}.
\label{eq:shift}
\end{equation}
The method then consists in building the \emph{Loewner} $\LL \in \Cplx^{m\times m}$ and \emph{shifted Loewner} $\sLL \in \Cplx^{m\times m}$ matrices defined as follows, for $i=1,\dots,m$ and $j=1,\dots,m$:
\begin{eq}
\begin{array}{rclcl}
[\LL]_{j,i} &=& \dfrac{\vv_j^H\rv_i - \lv_j^H\wv_i}{\mu_j - \lambda_i} 
&=& \dfrac{\lv_j^H\big( \Htran(\mu_j) - \Htran(\lambda_i) \big) \rv_i}{\mu_j - \lambda_i} \\
\,[\sLL]_{j,i} &=& \dfrac{\mu_j\vv_j^H\rv_i - \lambda_i\lv_j^H\wv_i}{\mu_j - \lambda_i}
&=& \dfrac{ \lv_j^H\big( \mu_j\Htran(\mu_j) - \lambda_i\Htran(\lambda_i) \big) \rv_i}{\mu_j - \lambda_i}
\end{array}.
\label{eq:loewnerMatrices}
\end{eq}
Then, the model $\Htran^m$ given by the following descriptor realisation,
\begin{equation}
\Hreal^m:\left \lbrace
\begin{array}{rcl}
\E^m  \delta\left\{\x(\cdot)\right\} &=& \A^m \x(\cdot) + \B^m \u(\cdot)\\
\y(\cdot) &=&\C^m \x(\cdot)
\end{array}
\right. ,
\label{eq:loewnerDescrR}
\end{equation}
where $\E^m = -\LL$, $\A^m = -\sLL$, $[\B^m]_k = \vv_k^H$ and $[\C^m]_k = \wv_k$ (for $k=1,\ldots,m)$, and whose related transfer function 
\begin{equation}
\Htran^m(\xi) = \C^m\pare{\mathbf \xi \E^m-\A^m}^{-1}\B^m,
\label{eq:loewnerDescrC}
\end{equation}
interpolates $\Htran$ at the given driving frequencies and directions defined in \cref{eq:loewnerInput}, \ie satisfies
\begin{equation}
\begin{array}{rcl}
\lv_j^H\Htran^m(\mu_j) &=& \lv_j^H\Htran(\mu_j) \\
\Htran^m(\lambda_i)\rv_i &= &\Htran(\lambda_i) \rv_i
\end{array}.
\label{eq:loewnerIntep}
\end{equation}

\begin{remark}[About $\delta\left\{\cdot\right\}$ and $\xi$ notations]
In \cref{eq:loewnerDescrR}, ``$(\cdot)$'' denotes the considered time-domain variable; this latter can either be ``$(t)$'' for continuous-time models ($t\in\Real_+$) or ``$[q]$'' for sampled-time models ($q\in\mathbb Z$). Similarly, in \cref{eq:loewnerDescrR} ``$\delta\left\{\cdot\right\}$'' stands as the shift operator being either $\delta\{\x(t)\}=\dx(t)$ in the continuous-time case and $\delta\{\x(q)\}=\x[q+1]$ in the sampled one. With reference to \cref{eq:loewnerDescrC} $\mathbf \xi$ stands as the associated complex version being the Laplace variable $\mathbf \xi=s$ in the continuous-time case and the forward shift $\mathbf \xi=z$ in the sampled-time one.
\end{remark}

Assuming that the number $2m$ of available data is large enough, then  it has been shown in \cite{Mayo:2007} that a minimal model $\Htran^n$ of dimension $n < m$ that still interpolates the data can be built with a projection of \cref{eq:loewnerDescrR} provided that, for $k=1,\ldots,2m$, where $z_k$ are as in \cref{eq:shift},
\begin{equation}
    \rank\pare{z_k \LL - \sLL} = \rank\pare{ [\LL,\sLL]} = \rank\pare{[\mathbb{L}^H,\sLL^H]^H} = n.
    \label{eq:rankCond}
\end{equation}
In that case, let us denote by $Y \in \Cplx^{m \times n}$ the matrix containing the first $n$ left singular vectors of $[\LL,\sLL]$ and $X \in \Cplx^{m \times n}$  the matrix containing the first $n$ right singular vectors of $[\LL^H,\sLL^H]^H$. Then,
\begin{equation}
    \E^n = Y^H \E^m X,\,
    \A^n = Y^H \A^m X,\,
    \B^n = Y^H \B^m,\,
    \C^n = \C^m X,
    \label{eq:proj}
\end{equation}
is a realisation of the model $\Htran^n$ given as, 
\begin{equation}
\Htran^n(\xi) = \C^n\pare{\mathbf \xi \E^n-\A^n}^{-1}\B^n,
\label{eq:loewnerDescrCn}
\end{equation}
with the same structure as \cref{eq:loewnerDescrC}, embedding a \emph{minimal McMillan degree} equal to $\rank\pare{\LL}$. The quadruple $\Hreal^n:(\E^n,\A^n,\B^n,\C^n)$ is a descriptor realisation of $\Htran^n$. Note that if $n$ in \cref{eq:rankCond} is greater than $\rank(\LL)$ then $\Htran^n$ can either have a direct-feedthrough or a polynomial part. 
Reader may note that the number $n$ of singular vectors composing $Y$ and $X$ used to project the system $\Htran^n$ in \cref{eq:proj} may be decreased even further than $n$ at the cost of approximate interpolation of the data. This allows for a trade-off between complexity of the resulting model and accuracy of the interpolation. Reader may note that Loewner extensions to some nonlinear systems exist, \eg bilinear \cite{Gosea:2016}. 

In this work we will always consider exact interpolation while the model reduction step, recalled in \cref{ssec:modred}, will be performed by (frequency limited) $\Htwo$-oriented interpolatory methods instead. These latter are detailed in a rather large framework in the following \cref{ssec:h2} and \cref{ssec:h2w}. 

\begin{remark}[Assumptions for model order reduction]
In what follows, we assume that $\Htran^n$ is stable, strictly proper and has semi-simple poles. These assumptions can be theoretically removed but at the cost of a more complicated developments, not appropriate in our gust load setting. Indeed, any polynomial or constant terms, as highlighted in \cref{eq:model1tf}, will be kept for simulation but discarded for control design, with a low if not null impact on the process.
\end{remark}

\subsection{Model dimension reduction}
\label{ssec:modred}

Given the finite $n$-th order function $\Htran^n$ equipped with a realisation defined by the quadruple $\Hreal^n:(\E^n,\A^n,\B^n,\C^n)$ given in \cref{eq:proj}, the model reduction goal consists in constructing reduced $r$-th order ($r\ll n$) model
\begin{eq}
\Hrealr: \left\{ 
\begin{array}{rcl} 
\Er\delta\left\{\xr(\cdot)\right\} &=&\Ar \xr(\cdot) +\Br \u(\cdot)  \\ 
\yr(\cdot) &=& \Cr \xr(\cdot) 
\end{array} \right. ,
\label{eq:redDescrR}
\end{eq}
where, $\xr(\cdot) \in \Real^{r}$ is the reduced internal variables and $\yr(\cdot)\in\Real^{n_y}$ is the approximated output, and where 
\begin{eq}
\Er \in \Real^{r \times r}, \Ar \in \Real^{r \times r}, \Br \in \Real^{r\times n_u} \text{ and }  \Cr \in \Real^{n_y \times r}
\end{eq}
are constant matrices, such that the input output behaviour of $\Htranr$ defined as
\begin{equation}
\Htranr(\xi)=\Cr\pare{\xi\Er-\Ar}^{-1}\Br,
\label{eq:redDescrC}
\end{equation}
is similar to $\Htran^n$, \ie for the same input $\u$, the mismatch $\y-\yr$ is small in some sense. 

\subsection{Rational $\Htwo$ approximation and reduction by interpolation}
\label{ssec:h2}

One way to find $\Htranr$ as in \cref{eq:redDescrC} is to solve the so-called $\Htwo$ model reduction problem\footnote{Optimising through the $\Htwo$ norm is relevant as a it provides an upper bound on the worst case time-domain error  in response to an input signal \cite{GugercinSIAM:2008}.}, which aims at solving the following problem:
\begin{equation}
\Htranr = \arg \min_{\Gtran \in \Htwo} \norm{\Htran^n-\Gtran}_{\Htwo}.
\label{eq:h2Optim}
\end{equation}
The most common approach to obtain the solution of \cref{eq:h2Optim} is to work with the \emph{first order necessary optimality conditions} which were developed in a series of papers, see \eg  \cite{Baratchart:1991,Spanos:1992} for theoretical insight (see also \cite{Baratchart:1986} for existence and general properties). The \emph{interpolation-based} approach revisited by a sequence of contributions \cite{Gallivan:2004,VanDooren:2008,GugercinSIAM:2008} and resulting in an interpolatory problem is used here. In the case of semi-simple poles only (the case of higher order poles is presented in \cite{VanDooren:2010}), if $\Htranr$, equipped with a realisation $\Hrealr:(\Er,\Ar,\Br,\Cr)$ is a solution of the $\Htwo$ approximation problem \cref{eq:h2Optim}, then
\begin{equation}
\begin{array}{rcl}
\Htran^n(\hat{\kappa}_l)\hat{\mathbf{b}}_l &=& \Htranr(\hat{\kappa}_l)\hat{\mathbf{b}}_l \\
\hat{\mathbf{c}}_l^H \Htran^n(\hat{\kappa}_l) &=&\hat{\mathbf{c}}_l^H \Htranr(\hat{\kappa}_l)\\
\hat{\mathbf{c}}_i^H \eval{\dfrac{d\Htran^n}{d\xi}}{\xi=\hat{\kappa}_l}\hat{\mathbf{b}}_l &=&\hat{\mathbf{c}}_l^H \eval{\dfrac{d\Htranr}{d\xi}}{\xi=\hat{\kappa}_l}\hat{\mathbf{b}}_l
\end{array},
\label{eq:h2Interp}
\end{equation}
where $[\hat{\bv}_1,\dots ,\hat{\bv}_r]^H=R\Br$ and $[\hat{\cv}_1,\dots ,\hat{\cv}_r]=\Cr L$ and where $L\in\Cplx^{r\times r}$ and $R\in\Cplx^{r\times r}$ are the left and right eigenvectors associated to $\hat{\lambda}_l$, the eigenvalues of the $(\Er,\Ar)$ pair. Then, in the continuous-time case, $\hat\kappa_l=-\hat\lambda_l$ while in the discrete-time one,  $\hat\kappa_l=1/\hat\lambda_l$ ($l=1,\dots ,r$) \cite{GugercinSIAM:2008,Bunse:2010}.

In \cite{GugercinSIAM:2008}, authors derive a Petrov-Galerkin like projection-based approach \cite{Villemagne:1987}, accompanied with a fixed point algorithm, celebrated as \mimo Iterative Rational Krylov Algorithm (\textbf{MIMO IRKA}), allowing reaching the optimality conditions \cref{eq:h2Interp}. The solution results in a procedure with a fairly cheap computational cost, embedding linear equation resolutions for which efficient methods are available \cite{Saad:1986,Lehoucq:1996,Sleijpen:1996,Higham:2002}. Details are skipped here but readers can also refer to the book \cite{AntoulasBook:2020} or monograph \cite[ Chap. 2]{PoussotHDR:2019} for practical details and numerical applications.

\subsection{Rational frequency-limited $\Htwo$ approximation and reduction by interpolation}
\label{ssec:h2w}

Similarly to the above $\Htwo$ model approximation problem, it might be interesting approximating over a frequency limited range $\Omega=[0,\omega]$ ($\omega\in\Real_+$) only. This is the purpose of the \emph{frequency-limited $\Htwo$}, shortly $\Htwow$, approximation problem defined as \cite{VuilleminH2ECC:2014,Petersson:2014}:
\begin{equation}
\Htranr = \arg \min_{\Gtran \in \Hinf} \norm{\Htran^n-\Gtran}_{\Htwow}.
\label{eq:h2wOptim}
\end{equation}
Interestingly, as in the seminal works of \cite{GugercinSIAM:2008} treating the $\Htwo$ objective, \cite[Chap. 8]{VuilleminPhd:2014} shown that this problem can also be recast as an interpolatory one. 

However, unlike the $\Htwo$ interpolation conditions, the frequency-limited ones do not involve directly the transfer functions $\Htran$ and $\Htranr$ but irrational functions $\mathbf T^n_\omega(\Htran)$ and $\mathbf{\hat T}_\omega(\Htranr)$, parametrised by  $\Htran$ and $\Htranr$. Here again, these functions should match at images of the poles of the reduced-order model (see details in \cite[Chap. 8]{VuilleminPhd:2014}). 

As it, these interpolatory conditions are difficult to practically exploit. Indeed, no Krylov-like subspace have been clearly identified yet. Consequently, a \mimo \irka like procedure as  in \cite{GugercinSIAM:2008} is not straightforward to develop\footnote{Note that in \cite{PontesSCL:2018}, a similar procedure achieving first order $\Htwo$ optimality with interpolatory conditions with an input-output delayed \lti model has been developed.}. So far, this $\Htwow$ problem has been attacked using the \darpo procedure \cite[Chap. 8]{VuilleminPhd:2014}, a descent algorithm, and \flistia procedure  \cite{VuilleminSSSC:2013}, involving frequency-limited gramian, taking advantage of a subset of the interpolation conditions \cref{eq:h2Interp}. 
In the remaining of the paper, we will consider the \flistia procedure developed in \cite{VuilleminSSSC:2013}.


%% file: sec-model.tex
Back to our GLA problem, we will now invoke the interpolatory results of \cref{sec:interp} to construct an appropriate control-oriented model for the GLA control design, simulation and analysis. The following \cref{sub:bizjetApprox} and \cref{sub:bizjetRed} successively describe the exact continuous-time rational approximation and frequency-limited model order reduction of the aeroelastic business jet aircraft models. The reason for these successive sections are justified in \cref{sub:bizjetLimit}. Numerical results are presented in \cref{sub:bizJetAppli}.

\subsection{Models limitations for control purpose}
\label{sub:bizjetLimit}

This BizJet use-case is described by models defined as \cref{eq:model1ss} in state-space form $\{\Hreal_i\}_{i=1}^{n_s}$ and \cref{eq:model1tf} in its transfer form $\{\Htran_i\}_{i=1}^{n_s}$. These relations are not appropriate to manipulate in a control design objective for the following reasons:  \emph{(i)} the internal vector dimension of each model $\Hreal^{(i)}$ is large (\ie $N_i\approx 300$), leading to medium-scale matrices and a computational burden inappropriate with the control design; \emph{(ii)} the presence of internal delays renders the model irrational and results in an infinite number of eigenvalues, for which dedicated tools exist but are complex to manipulate in an industrial context; \emph{(iii)} the $\E^{(i)}$ matrix being rank deflective, the computational tools are not always appropriate and transferable in an applicative context. 

\subsection{BizJet aircraft rational and polynomial approximation}
\label{sub:bizjetApprox}

The first step of the process consists in replacing each infinite\footnote{These models, embedding a state-space of order $N_i$ are of infinite dimension due to the internal delays, leading to infinite number of eigenvalues.} dimensional model $\{\Htran_i\}_{i=1}^{n_s}$ \cref{eq:model1tf}  by a finite order rational $n_i$-th order one, $\{\Htran_i^{n_i}\}_{i=1}^{n_s}$ ($i=1,\dots n_s$). 
By considering each $n_s$ continuous-time model of the considered GLA use-case described in \cref{sec:intro}, one can apply the rational approximation by interpolation using the method recalled in \cref{ssec:loe}.  Then one obtains a set of $n_s$ continuous-time rational approximated models $\{\Htran_i^{n_i}\}_{i=1}^{n_s}$ as  \cref{eq:loewnerDescrCn} equipped with realisation $\{\Hreal_i^{n_i}\}_{i=1}^{n_s}$ with matrices as \cref{eq:proj}. Note that each of the $n_s$ models may have a different dimension $n_i$; this latter being automatically computed by the rank condition given in \cref{eq:rankCond}. Here, as the original models \cref{eq:model1tf} are given in continuous-time, it is convenient to select the interpolation points along the imaginary axis. In our case, for all configurations, the interpolating points \cref{eq:shift} have been selected as follows : 
\begin{equation}
\left\{z_k\right\}_{k=1}^{2m} =
\underbrace{\left\{ \imath \omega_j,-\imath \omega_j \right\}_{j=1}^{m/2}}_{\left\{\mu_j\right\}_{j=1}^m}
\cup
\underbrace{\left\{ \imath \nu_i,-\imath \nu_i \right\}_{i= 1}^{m/2}}_{\left\{\lambda_i\right\}_{i=1}^m},
\end{equation}
where $\omega_j,\nu_i \in \Real_+$ are the pulsations at which one evaluates each transfer $\{\Htran_i\}_{i=1}^{n_s}$, with $m=1000$. In our application $\omega_j$ and $\nu_i$ are selected to be logarithmically spaced from $10^{-2}$ to $10^3$. This choice allows to focus on the frequency range of interest. Indeed, in the case of irrational models, the method is efficient for interpolation but not necessary for extrapolation. 

In addition, as the stability of the obtained rational model $\{\Htran_i^{n_i}\}_{i=1}^{n_s}$ is not guaranteed by the Loewner interpolatory procedure, a \emph{post stabilisation} is performed using the procedure presented in \cite{Kohler:2014}. This latter consists in projecting the rational models $\{\Htran_i^{n_i}\}_{i=1}^{n_s}$ onto its closest stable subset, here using the $\Hinf$-norm, leading to a stable model of the same dimension. Mathematically, given a realisation $\Hreal$ associated to $\Htran\in \Linf$, one aims at finding $P_\infty(\Htran) \in \Hinf$ such that,
\begin{equation}
P_\infty(\Htran) = \arg \inf_{\mathbf G \in \Hinf}\norm{\Htran-\mathbf G}_{\mathcal H_\infty}.
\label{eq:stableApprox}
\end{equation}
Technical details and assumption can be found in \cite{Kohler:2014}. Each model now share a rational structure as the one given in \cref{eq:redDescrR}-\cref{eq:redDescrC}, and is now stable and of finite order $n_i$, for $i=1,\dots , n_s$. Note that here, the stability enforcement is performed since one knows that original models are  structurally and physically stable. At this point, the delay difficulty is removed and traded with rational order model $n_i$ automatically determined by \cref{eq:rankCond}. 

\begin{remark}[About a Pad\'e delay approximation]
It can be appealing to replace the delays with a Pad\'e approximation, which preserves the gain but modifies the phase. Indeed, while this is classically performed in many applications, it is, to the authors experience, not the best way to deal with delays, especially in flexible structure models where phase vary a lot. In addition, the use of Pad\'e will drastically increases the model internal vector. Although it may not be a problem, Pad\'e  usually results in poor accuracy, specifically in flexible structure, affecting the phase. Instead, we prefer rational multi-point interpolation while Pad\'e may be viewed as an interpolation at higher order, at zero. 
\end{remark}

\subsection{BizJet aircraft control-oriented model reduction}
\label{sub:bizjetRed}

As a second step, and rooted on $\{\Htran_i^{n_i}\}_{i=1}^{n_s}$, the obtained continuous-time rational models, one now invokes again the \emph{interpolatory framework} for a dimension reduction. Now we follow the approach detailed in  \cref{ssec:h2w} to reduce the model over a frequency-limited range. For the considered application, and in view of control design, it is important to reduce as much as possible the model while staying representative. Indeed, as the controller designed later is a solution of an optimisation problem, the simpler the model is, the more accurate and faster the optimisation is.

In addition, as the GLA function should act in a given  frequency range without altering the flight control laws, it is relevant to reduce the model over a frequency limited range $[0,\omega]$, only (where $\omega\in\Real_+$). \flistia process \cite{VuilleminSSSC:2013} being also a fixed point procedure, its initialisation is done by selecting starting interpolating points as follows:
\begin{equation}
\begin{array}{rcll}
\{\hat{\kappa}^{(0)}\}_{k=1}^{r} &=& \{\imath \omega_k,-\imath\omega_k\}_{k=1}^{r/2} & \text{ if $r$ is even}\\
\{\hat{\kappa}^{(0)}\}_{k=1}^{r} &=& \{\imath \omega_k,-\imath\omega_k\}_{k=1}^{\left \lfloor{r/2}\right \rfloor} \cup \omega_0 & \text{ if $r$ is odd}
\end{array},
\end{equation}
where $0<\omega_k<\omega$ and $\omega_0<\omega$. The interpolatory conditions presented in \cref{eq:h2Interp} can also be partially ensured.

\subsection{Application to the BizJet models}
\label{sub:bizJetAppli}

\begin{figure}[htbp]
  \centering
  \includegraphics[width=.6\textwidth]{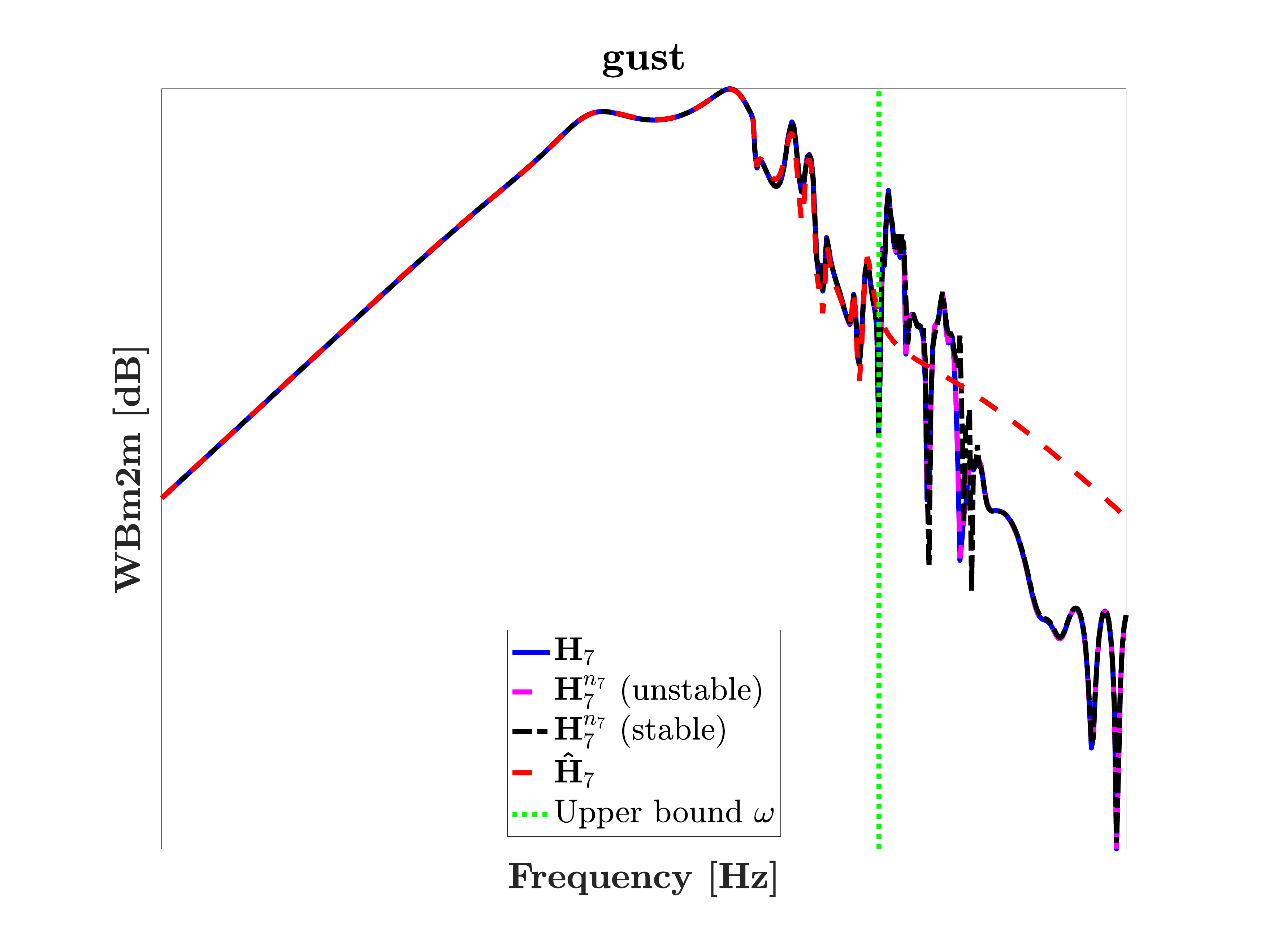}
  \includegraphics[width=.6\textwidth]{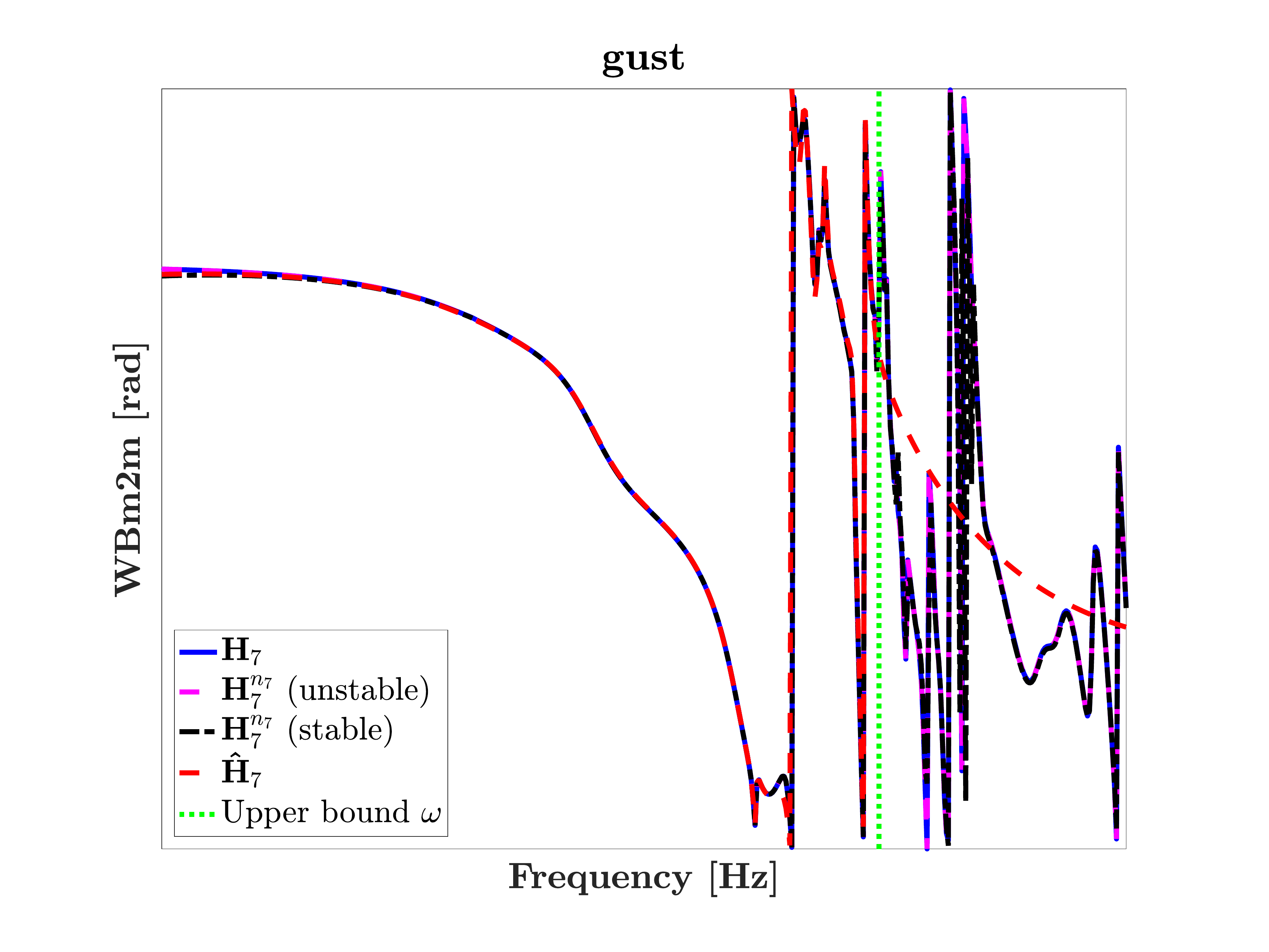}
  \caption{Gain (top) and phase (bottom) frequency responses of one transfer from gust to a wing root bending moment. The original model $\Htran_7$ (solid blue), the approximated and approximated plus projected stable models $\Htran^{n_7}_7$ (dashed magenta dash dotted black, respectively) and the reduced order model $\Htranr_7$ with dimension $r=30$. Upper bound of the reduction frequency band is materialised by the vertical green dashed line.}
  \label{fig:bizjetApproxBode}
\end{figure}

\begin{figure}[htbp]
  \centering
  \includegraphics[width=.6\textwidth]{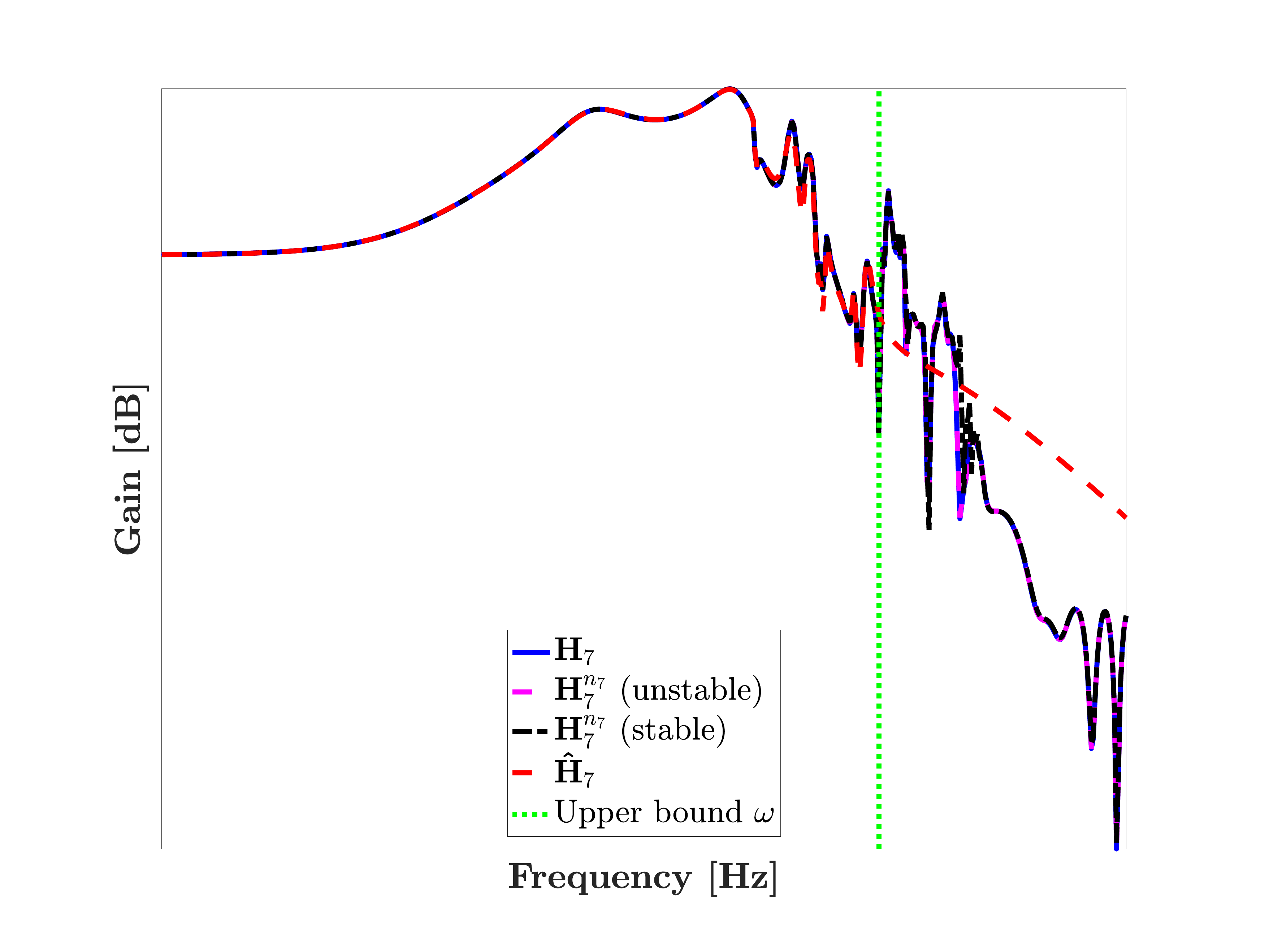}
  \includegraphics[width=.6\textwidth]{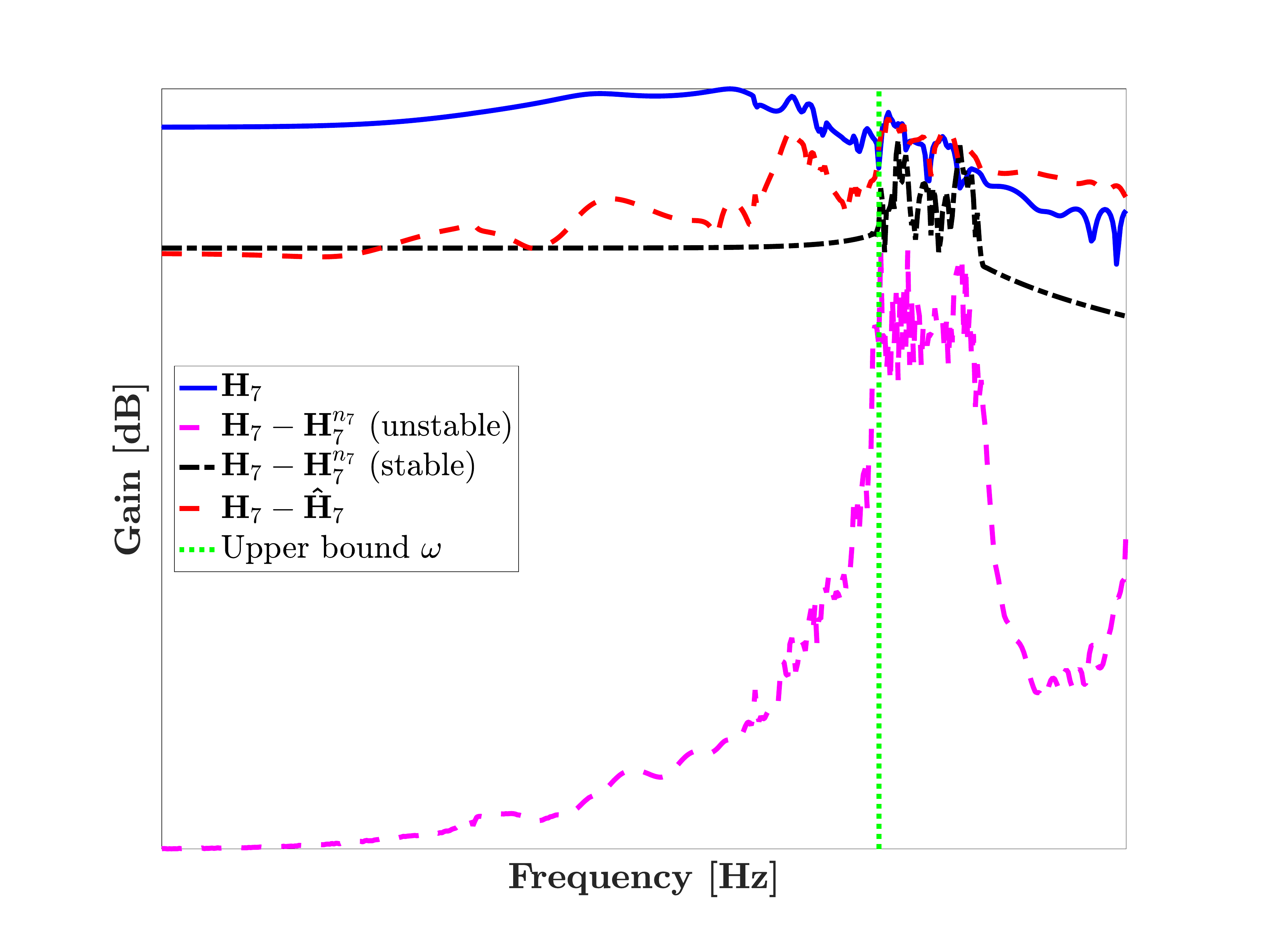}
  \caption{Gain (top) and gain error (bottom) of the singular frequency responses of the different models.}
  \label{fig:bizjetApproxSigma}
\end{figure}

As an illustration, \cref{fig:bizjetApproxBode} compares the frequency and phase responses from the gust input to a wing root bending moment of the original irrational model $\Htran_7$ with its rational approximations $\Htran_7^{n_7}$ and frequency-limited reduced model $\Htranr_7$ for one single configuration point. Sigma plot and mismatch errors are also reported in \cref{fig:bizjetApproxSigma}. All other cases are similarly obtained. 

With reference to \cref{fig:bizjetApproxBode} and \cref{fig:bizjetApproxSigma}, the following comments can be done. First, the rational approximation, satisfying the tangential interpolatory conditions \cref{eq:loewnerIntep}, allows to accurately capture the descriptor delayed model \cref{eq:model1tf} gain, phase and sigma, with a simple rational model (here for the 7-th use-case, $n_7=211$) of the form \cref{eq:loewnerDescrCn}. Note that the projection \cite{Kohler:2014} performed afterward to force the model to be stable let the matching accurate enough. Second, the frequency-limited order reduction still preserves a good matching in the considered frequency range, with a rational model as \cref{eq:redDescrC} of dimension $r=30$ only. Analysing the bottom frame of \cref{fig:bizjetApproxSigma}, the $\Htran_7-\Htran_7^{n7}$ (unstable) mismatch is close to machine precision up to the upper frequency limit $\omega$; this is caused by the tangential directions which may produce less accurate transfers. When projecting to obtain a stable model, the $\Htran_7-\Htran_7^{n7}$ (stable) mismatch  results in a loss of accuracy, traded with the stability property. The reduced model  $\Htranr_7$ clearly is accurate below the upper frequency bound, and can be considered as simple but accurate enough for GLA control design.

%% file: sec-control.tex
Given the continuous-time reduced order models $\{\Htranr_i\}_{i=1}^{n_s}$, we are now ready to design the GLA control function. As this control law has to be implemented on a on-board computer sampled with a fixed step time $h\in \Real_+$, a sampled-time (or discrete-time) controller is sought. In the rest of this section, we first present the context of discrete-time control design in \cref{ssec:intro}. Then, based on the continuous-time reduced models $\{\Htranr_i\}_{i=1}^{n_s}$,  the continuous-time synthesis of $\mathbf K$, in the continuous-time $\Hinf$-norm, is briefly exposed in \cref{ssec:continousK}. Then, \cref{ssec:d_pb} exposes the discretisation problem, which is then solved in the \emph{new interpolatory framework} in \cref{ssec:discLoe}, leading to \cref{theAlg}.

\subsection{Preliminary words on discrete and continuous-time control design}
\label{ssec:intro}

\begin{figure}
    \centering
      \input{./figures/sd-synth}
    \caption{Different paths to design a discrete control-law $\mathbf K_d$ from an analog plant $\Htran$.}
    \label{fig:synth-scheme}
\end{figure}
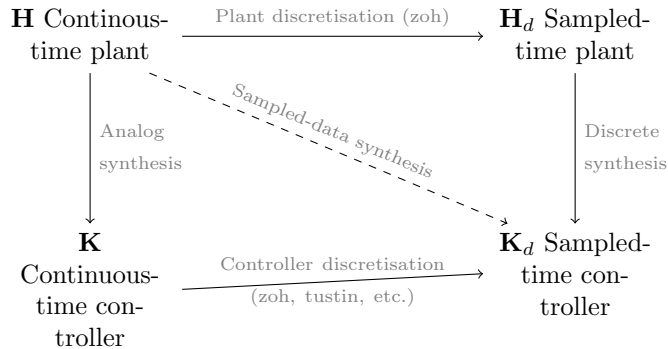

In dynamical systems and control theory, the continuous-time and discrete-time domains coexist. While most of the tools available in one domain have a counterpart in the other, it is not unusual that a specific application requires to switch from one domain to the other. In particular, in the considered GLA control-oriented application, engineers start from a  continuous-time physical model set  $\{\Htran_i\}_{i=1}^{n_s}$, simplified by a reduced continuous-time one $\{\Htranr_i\}_{i=1}^{n_s}$. The latter is then used for the design of the GLA control law. To this aim, as illustrated schematically in \cref{fig:synth-scheme}, three approaches can be conducted: \emph{(i)} one may discretise the analog plants $\{\Htran_i\}_{i=1}^{n_s}$ and then design a discrete control-law $\mathbf K_d$ with the same sampling period $h$, \emph{(ii)} conversely, one may design an analog control-law $\mathbf K$ and then discretise it, \emph{(iii)} or, using dedicated techniques from sampled-data systems theory (see \eg \cite[Chap. 12-13]{chen1995optimal}), one may directly synthesise $\mathbf K_d$ from $\{\Htran_i\}_{i=1}^{n_s}$.

While the direct nature of the latter method \emph{(iii)} is appealing, it requires dedicated theoretical and numerical tools that are not as widespread as usual ones, especially in the industry where this approach would require their whole control design and analysis process to be rethought. For these reasons, one focuses here on the indirect methods that remain of practical interest, and more specifically on the continuous controller, followed by a discretisation. Performing in the other ways, \ie discretising the model and then synthesising the controller would lead to an inaccurate model response due to the low frequency sampling imposed by the on-board computer.

For both indirect approaches, a discretisation step is required. This may have a detrimental impact with respect to the expected dynamical behaviour, especially if hard computational constraints on the sampling period $h$ are imposed by the technology, which may happen with critical systems. In that context, the availability of an efficient discretisation method is of particular interest. Here, an approach based on the \emph{Loewner interpolatory framework} \cite{Mayo:2007}, recalled in \cref{ssec:loe}, is used. It offers an interesting alternative to usual discretisation processes as it enables to reach a better frequency and time-domain matching with the continuous-time description. Before detailing this approach, for sake of completeness, \cref{ssec:continousK} briefly presents the continuous-time GLA controller design.

\subsection{Preliminary continuous-time GLA controller design}
\label{ssec:continousK}

In the considered application, as the objective is to attenuate a worst case amplification, the $\Hinf$-norm minimisation framework is clearly tailored. This optimisation method is then used, on the basis of the reduced continuous-time model representation. Note that this framework, and linear ones, are largely used and mastered in aeronautical applications \cite{OssmannEuroGNC:2019,Pusch:2018,Alama:2015}, and is preferred here also. Being given the reduced continuous-time  models $\{\Htranr_i\}_{i=1}^{n_s}$, the continuous-time controller $\mathbf K$, mapping outputs $\y$ to control inputs $\u$, as interconnected on \cref{fig:global}, is obtained by solving the following $\Hinf$-norm problem, for $i=1,\dots n_s$,
\begin{eq}
{\mathbf K} = \arg \min_{\tilde{\mathbf K}\in \mathcal K \subseteq \Htwo} \norm{\mathcal F_l(\mathbf G(\Htranr_i) , \tilde{\mathbf K})}_{\mathcal H_{\infty}},
\label{eq:controlHinf}
\end{eq}
where $\mathbf{G}(\Htranr_i)\in\Hinf$ is the so-called generalised plant involving the input and output weight functions (not detailed here as out of the scope) and the reduced order models. Moreover $\mathcal F_l(\cdot,\cdot)$ denotes the lower fractional operator. The control objective thus consists in finding the  locally-optimal continuous-time controller $\mathbf K$, such that \cref{eq:controlHinf} is solved, \ie such that the $\Hinf$-norm of the $\mathcal F_l(\mathbf G(\Htranr_i) , {\mathbf K})$  loop, mapping exogenous inputs $\mathbf w$ to performance output $\mathbf z$ (here being the loads and some tracking objectives), is minimised. In the considered case, one seeks this controller such that it belongs to a subset $\mathcal K \subseteq \Htwo$, reducing the controller search space to reduced order structures without direct feed-through, avoiding issues for industrial implementation. The problem is solved by the routine developed by the authors of \cite{Apkarian:2006} and results in a stable moderate $n_c$-th order \lti continuous-time controller $\mathbf K$.

\subsection{Controller discretisation problem statement and discretisation error derivation}
\label{ssec:d_pb}

Being given the linear continuous-time controller $\mathbf K$, the objective here is to determine, for a fixed time-step $h \in \Real_+$, a discrete-time controller represented by the recurrence state-space equation,
\begin{equation}
\mathcal C_d: \left\{ 
    \begin{array}{rcl}
         \x_d[q+1]&=& \A_d \x_d[q] + \B_d \y_d[q]  \\
         \u_d[q]& = & \C_d \x_d[q] + \D_d \y_d[q]
    \end{array}
   \right.
    \label{eq:sysd}
\end{equation}
where $\x_d[q] \in \Real^{n_d}$, $\u_d[q] \in \Real^{n_u}$, $\y_d[q] \in \Real^{n_y}$ are the internal variables, sampled control and measurements signals respectively. One seeks for \cref{eq:sysd} such that \emph{(i)} its associated transfer function $\mathbf K_d(z) = \C_d(zI_{n_d} - \A_d)^{-1} \B_d + \D_d$ is stable, \ie $\mathbf K_d \in \dhinf$, and \emph{(ii)} the input-output behaviour of $\mathbf K$ is well reproduced by $\mathbf K_d$ up to the Nyquist frequency. While discretisation methods such as the bilinear (Tustin) are clearly able to build a discrete-time model satisfying \emph{(i)} the input-output behaviour may be quite far from the original one when $h$ is too large. Here, a method is proposed to build such a model by first using the \emph{Loewner framework} to interpolate a specific set of frequency data and then, similarly to \cref{sub:bizjetApprox}, projecting the resulting model onto the stable subspace $\dhinf$ to enforce sampled-time controller stability.


\subsubsection{Measure of the discretisation error}
\label{sssec:discErr}

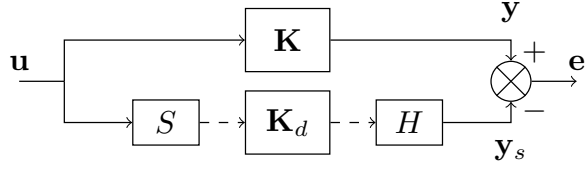
\begin{figure}
\centering
\scalebox{1.2}{
\begin{tikzpicture}
  \node (G) [anchor = west, draw, minimum height= 0.7cm, text width = 0.7cm, text centered ]{$\mathbf K$};
  \node (gd) at ($(G.south)+(0,-0.2cm)$) [anchor = north, draw,minimum height= 0.7cm, text width = 0.7cm, text centered ] {$\mathbf K_d$};
  \node (st) at ($(gd.west) + (-0.5cm,0)$) [anchor = east, draw,minimum height= 0.5cm, text width = 0.5cm, text centered ] {$S$};
  \node (ht) at ($(gd.east) + (0.5cm,0)$) [anchor = west, draw,minimum height= 0.5cm, text width = 0.5cm, text centered ] {$H$};

  \node (add) [draw, circle, text width=4pt] at ($0.5*(G.east) + 0.5*(gd.east) + (2cm,0)$) {};

 \node (start) at ($0.5*(G.west) + 0.5*(gd.west) - (2cm,0)$) {};
  \draw [->] (start.center) |- (G.west);
  \draw [->] (start.center) |- (st.west);
  \draw [->] (G.east) -| (add.north) ;
 \draw [->] (ht.east) -| (add.south) ;
  \draw [dashed, ->] (st.east) -- (gd.west);
  \draw [dashed, ->] (gd.east) -- (ht.west);
  \draw [-] (start.center) --+ (-0.5cm,0) node [above] {$\u$};
  \draw [->] (add.east) --+ (0.5cm,0) node [above] {$\mathbf e$};
  \node [anchor=west] at ($(add.south)+(0,-0.1cm)$) {$-$};
        \node [anchor=west] at ($(add.north)+(0,0.1cm)$) {$+$};

  \node [anchor=north] at ($(add.south)+(0,-0.3cm)$) {$\y_s$};
  \node [anchor=south] at ($(add.north)+(0,0.3cm)$) {$\y$};

  \draw (add.45)--(add.225);
  \draw (add.135) -- (add.315);
\end{tikzpicture}
}
\caption{Interconnection for the measurement of the discretisation error between $G$ and $G_d$.}
\label{fig:discretisation_error}
\end{figure}

Quantifying the error induced by the discretisation of $\mathbf K$ is not trivial due to the incompatible nature with $\mathbf K_d$ that prevents from interconnecting the two systems. Indeed, one cannot just consider $\mathbf K-\mathbf K_d$.  Digital to analog converters are required. For a sampling time $h\in\Real_+$, the latter are modelled here as the ideal sampler $S$ and holder $H$. Then, the error system is given by the interconnection of  \cref{fig:discretisation_error} where discrete-time signals are represented by dashed lines. Such an interconnection of continuous and discrete time models is called a \emph{Sampled-data System} (\textbf{SD}).

The controller $\tilde{\mathbf K} = S \mathbf K_d H$ is no longer \lti but $h$-periodic and consequently, so is the error $\mathbf K - \tilde{\mathbf K}$ for which usual system norms cannot directly be applied.  This problem has been addressed in the literature for direct \textbf{SD} synthesis (see \eg \cite{bamieh:lifting:1991,chen1995optimal} and references therein). The recurring idea is to use continuous lifting to transform a periodic system into a discrete-time \lti model with infinite input and output spaces on which equivalent $\Htwo$ or $\Hinf$ norms can be defined.

While this framework appears well suited to evaluate $\mathbf K - \tilde{\mathbf K}$, the infinite dimensionality of the lifted model makes it quite technical. That is why here, a more straightforward approach based on the \emph{frequency characterisation of the discretisation error} formulated in \cite[section 3.5]{chen1995optimal} is considered instead. For that purpose, models for the ideal sampler and holder are recalled in \cref{sssec:idealsh} and the frequency error is presented in \cref{sssec:freqerr}\footnote{Note that in \cite{chen1995optimal}, the authors use the $\lambda$-transform while here, one considers the $z$-transform, the sign of the exponential in the Fourier transforms of discrete signal is thus modified.}.

\subsubsection{Ideal sampler and holder}
\label{sssec:idealsh}

Let us consider a continuous-time signal $v(t)$ and the sampling period $h$. The ideal sampler $S$ transforms $v(t)$ into a discrete sequence $v_d[q]$ such that,
\begin{equation}
    v_d[q] = v(qh),\, q\in\mathbb{Z}.
\end{equation}
As shown in \cite[Lemma 3.3.1]{chen1995optimal}, the Fourier transforms $\overline{v}_d =\mathcal{F}(v_d)$ and $\overline{v} = \mathcal{F}(v)$ are linked as follows,
\begin{equation}
    \overline{v}_d(e^{\imath\omega h}) = \frac{1}{h} \sum_{q\in\mathbb{Z}} \overline{v}(\imath \omega + \imath q \omega_s),
    \label{eq:freq_sampler}
\end{equation}
where $\omega_s = 2\pi/h$ is the sampling frequency. Equation \cref{eq:freq_sampler} highlights the frequency aliasing phenomena since all the multiples of the sampling frequency are indistinguishable in the output. Note that the sampling operator $S$ is not bounded for any signal in $\mathcal{L}_2(\mathbb{R})$ as shown in \cite{chen1991input}. To be bounded, the input signal $v$ must be restricted to the class of band-limited $\mathcal{L}_2 (\mathbb{R})$ signals or it must be filtered by a finite-dimensional stable and strictly causal system, \ie an anti-aliasing filter.

Similarly, the holder $H$ transforms a sequence $v_d[q]$ into a continuous-time signal $v(t)$ such that,
\begin{equation}
    v(t) = v_d[q],\quad qh\leq t < (q+1)h.
\end{equation}
The impulse response of the holder can be defined as the difference of two unit steps delayed by $h$. Let 
\begin{equation}
R(s) = \frac{1 - e^{-sh}}{sh}
\end{equation}
be the associated transfer function. Then, as shown in \cite[Lemma 3.3.2]{chen1995optimal}, the Fourier transforms of $v$ and $v_d$ are linked as follows,
\begin{equation}
    \overline{v}(\imath \omega) = h R(\imath\omega) \overline{v}_d(e^{\imath\omega h}).
    \label{eq:freq_holder}
\end{equation}

Back to \cref{fig:discretisation_error}, coupling equations \eqref{eq:freq_sampler} and \eqref{eq:freq_holder} enables to express the frequency-domain relationship between $\u$ and $\y_s$ (for $\abs{\omega}<\abs{\omega_s}$),
\begin{equation}
    \overline{\y}_s(\imath\omega) =  R(\imath\omega) \mathbf K_d(e^{\imath\omega h}) \sum_{k\in \mathbb{Z}} \overline{\u}(\imath\omega + \imath q \omega_s).
    \label{eq:freq_relation}
\end{equation}

\subsubsection{A frequency domain error}
\label{sssec:freqerr}

Using equation \eqref{eq:freq_relation}, one can express the frequency-domain relationship between $\u$ and the discretisation error $\mathbf e = \y - \y_s$ from Figure \ref{fig:discretisation_error} as
\begin{equation}
    \overline{\mathbf e}(\imath\omega) = \mathbf K(\imath\omega) \overline{\u}(\imath\omega) - R(\imath\omega) \mathbf K_d(e^{\imath\omega h})\sum_{k\in\mathbb{Z}} \overline{\u}(\imath\omega + \imath q\omega_s).
    \label{eq:freq_err}
\end{equation}
Assuming that $\u$ is band-limited, \ie that $\overline{\u}(\imath\omega) = 0$ for $|\omega| > \omega_s/2 = \omega_N$, then \eqref{eq:freq_err} becomes
\begin{equation}
    \overline{\mathbf e}(\imath\omega) = \left ( \mathbf K(\imath\omega) - R(\imath\omega) \mathbf K_d(e^{\imath \omega h}) \right ) \overline{\u}(\imath\omega) .
    \label{eq:simp_freq_err}
\end{equation}
This readily suggests to consider
\begin{equation}
    e_{\infty}(\mathbf K,\mathbf K_d) =  \max_{\omega < \omega_N} \left | \mathbf K(\imath\omega) - R(\imath\omega) \mathbf K_d(e^{\imath\omega h}) \right |,
    \label{eq:err}
\end{equation}
to quantify the discretisation error. This frequency-domain characterisation of the error inspired the discretisation process presented in the next section. For the \mimo case, the absolute value in \eqref{eq:err} may be replaced by the $2$-norm.

\subsection{Application of the Loewner framework for discretisation}
\label{ssec:discLoe}

Now we are ready to describe the \emph{interpolatory driven discretisation process} allowing to get $\mathbf K_d$ from the continuous-time controller $\mathbf K$.

\subsubsection{Principle}

Being given the considerations of \cref{ssec:d_pb}, let us consider the transfer function of the controller $\mathbf K \in \Hinf$ to be discretised at the sampling time $h$. Building a $n_c$-th order discrete-time model $\mathbf K_d$ that matches the input-output behaviour of $\mathbf K$, the frequency-domain characterisation of the discretisation error \cref{eq:err} suggests that $\mathbf K_d$ must be such that
\begin{equation}
    R(\imath\omega) \mathbf K_d(e^{\imath\omega h}) = \mathbf K(\imath\omega),
    \label{eq:interp_cond}
\end{equation}
for $|\omega| < \omega_N$. Equation \cref{eq:interp_cond} represents an infinite number of interpolation conditions that may be approximated by sampling the interval $[0,\omega_N]$ such that for $k=1,\ldots,2m$,
\begin{equation}
    R(\imath\omega_k) \mathbf K_d(e^{\imath\omega_k h}) = \mathbf K(\imath\omega_k).
    \label{eq:sampled_interp_cond}
\end{equation}
Such a model can be built by applying the \emph{Loewner interpolation} framework recalled in \cref{ssec:loe} to the following set of frequency data,
\begin{equation}
    \left \lbrace e^{\imath\omega_k h}, R(\imath\omega_k)^{-1} \mathbf K(\imath\omega_k) \right \rbrace_{k=1}^{2m},
    \label{eq:dataset}
\end{equation}
where $e^{\imath\omega_k h}$ are the new interpolating points and $R(\imath\omega_k)^{-1} \mathbf K(\imath\omega_k)$ the new function to evaluate. With reference to \cref{eq:loewnerInput} and \cref{eq:shift} $e^{\imath\omega_k h}$  are the $z_k$. Then by splitting them in the same ways, $R(\imath\omega_k)^{-1} \mathbf K(\imath\omega_k)$ represents the couple $\vv_j^H$ and $\wv_i$. Following the notations of \cref{ssec:loe}, one  obtains the $m$-th order $\mathbf K_d^m$ controller that can be lowered without loss of interpolatory accuracy to $\mathbf K_d^n$ ($n<m$). Obviously, should the order $n_c$ be lower than the McMillan order $n$ of the exact interpolating model given by the Loewner approach, then the interpolation conditions \eqref{eq:sampled_interp_cond} will not be perfectly satisfied, but the controller will keep its original dimension. 

\subsubsection{About the stability of the discretised model}
\label{sssec:stab}

The Loewner framework does not ensure the stability of the resulting interpolaed model, here controller. Therefore, even if $\mathbf K \in \Htwo$ (or $\Hinf$), the transfer  function $\mathbf K_d$ obtained via the process described  above may not lie in $\dhinf$ which is a major drawback in comparison to stability preserving discretisation schemes like the bilinear (Tustin) of backward ones.

To overcome this issue, one can apply the same process as in \cite{gosea:stability:2016} and as performed in \cref{sub:bizjetApprox} (in the continuous-time). It consists in projecting the unstable model $\mathbf K_d$ onto $\dhinf$ so that the $\mathcal{L}_\infty$ norm between $\mathbf K_d$ and its projection is minimised. This is the so-called Nehari problem for which solutions have been given in the continuous-time domain \cite{Glover:1984} and in the discrete-time domain, see \eg \cite{mari:modifications:2000}.

Let us denote by $P_\infty$ this projection operator so that if $\mathbf K_d \in \mathcal{L}_\infty(\ucir) $ then $P_\infty(\mathbf K_d) \in \dhinf$ minimises 
\begin{equation}
\Vert \mathbf K_d - P_\infty(\mathbf K_d) \Vert_{\mathcal{L}_\infty}. 
\end{equation}
Note that in the sampled-time case, the order of $P_\infty(\mathbf K_d)$ depends on the number of unstable poles of $\mathbf K_d$ and is lower than $n_c$ when $\mathbf K_d$ is unstable. In particular, if $\mathbf K_d$ has $n_+$ stable and $n_-$ unstable poles and that $m$ is the multiplicity of the largest unstable Hankel singular value, then $P_\infty(\mathbf K_d)$ has order $n_+ + n_- - m$ (see \cite{mari:modifications:2000}). To avoid this issue and for a numerically more robust approach, sub-optimal projection methods may alternatively be considered (see \eg \cite{Kohler:2014}).



\begin{remark}[Alternative approach]
\label{rq:ldproj}
A similar problem can be considered using the $\mathcal{L}_2$-norm and is actually much easier to solve considering the decomposition $\mathcal{L}_2(\ucir) = \mathcal{H}_2(\udisk) \oplus \mathcal{H}_2(\udiskcomp)$. The solution is simply obtained by discarding the unstable part of the model. However this solution has generally a greater impact on the frequency behaviour of the model which is not desirable here.
\end{remark}

\subsubsection{Loewner-driven discretisation}

\begin{algorithm}[t]
\caption{Loewner-driven discretisation} \label{theAlg}
\begin{algorithmic}[1]
\REQUIRE A continuous-time model $\mathbf K \in \chinf$, a sampling time $h>0$, an upper bound $\bar{n}$ of the desired order and a number $m$ of interpolation points.
\STATE Sample the interval $]0,\omega_N[$ in $2m$ points $\omega_k$
\STATE Evaluate $R(\imath\omega_k)^{-1} \mathbf K(\imath\omega_k)$
\STATE Apply Loewner to the data-set \eqref{eq:dataset} to get a first model $\mathbf K_d^m$ that matches all the data and get the underlying minimal McMillian order $n$
\STATE Set $r=\min(n,\bar{n})$
\STATE Reduce $\mathbf K_d^n$ to $\mathbf K_d^r$ using any interpolation method 
\STATE Project $\mathbf K_d^r$ onto a stable subspace, \ie compute the $n_c$-th order $\mathbf K_d = P_\infty(\mathbf K_d^r)$
\RETURN the model $\mathbf K_d\in\dhinf$ of order $n_c\leq r \leq \bar{n}$
\end{algorithmic}
\end{algorithm}

The  Loewner-driven discretisation process of the controller is summarised in \cref{theAlg}. The following comments can be added:
\begin{itemize}
    \item the reduction step from $\mathbf K_d^n$ to $\mathbf K_d^r$ (step $5$) may also be achieved by standard model approximation techniques as the one presented in \cref{ssec:h2} or, more interestingly, using the frequency-limited version of \cref{ssec:h2w}. 
Then $\mathbf K_d^r$ should be projected on $\dhinf$ and finally a stability preserving model reduction method could be used to obtain $\mathbf K_d$.
    \item As both $\mathbf K_d^r$ and $\mathbf K_d$ are easily available during the process, $\Vert \mathbf K_d^r - \mathbf K_d \Vert_{\mathcal{L}_\infty}$ can be evaluated to get an estimation of the discretisation error $e_\infty(\mathbf K,\mathbf K_d)$ in \eqref{eq:err}. \item Consequently, when $n_c = r$, computing $\Vert \mathbf K_d^r - \mathbf K_d \Vert_{\mathcal{L}_\infty}$ gives insight on whether the maximal allowed order $\bar{n}$ is enough to ensure a low error.
    \item If $\mathbf K_d^r$ is unstable, then step $6$ leads to a loss of order (see section \cref{sssec:stab}). Therefore, when $n_c < r$ (which is likely to happen\footnote{Indeed, one tries to approximate data coming from an infinite dimensional model by a rational one. A large order $r$ is generally required to achieve an exact interpolation.}), $r$ should be increased above $\bar{n}$ so that $P_\infty(\mathbf K_d)$ is of order $\bar{n}$ thus avoiding any loss of accuracy.
\end{itemize}

\begin{figure}
    \centering
    \includegraphics[width=0.8\linewidth]{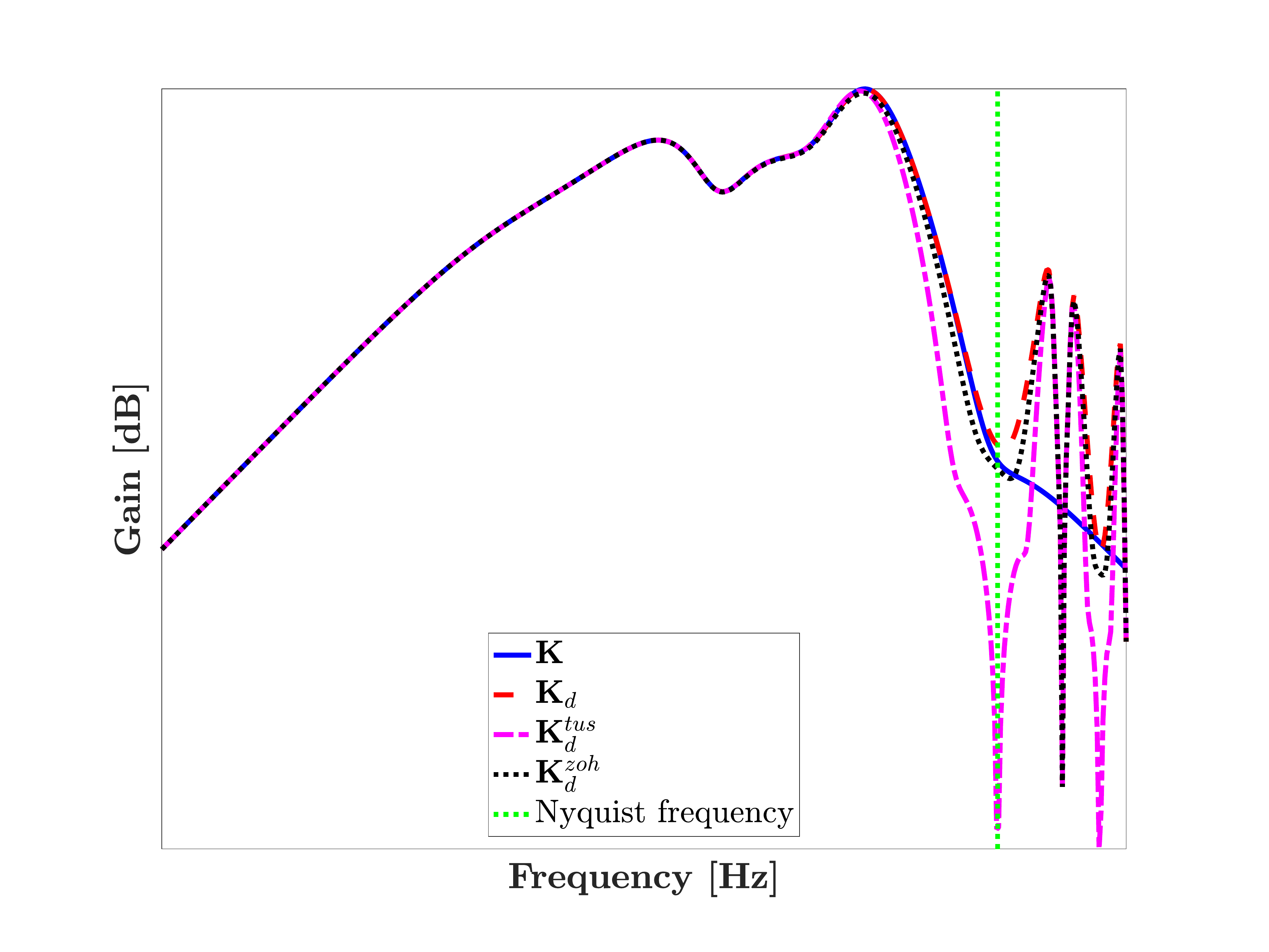}
  \caption{Comparison of the sigma plot of different controllers: the continuous-time one $\mathbf K$ with holder (solid blue), the discrete-time obtained with backward method $\mathbf K_d^{zoh}$ (dotted black), the discrete-time obtained with bilinear method $\mathbf K_d^{tus}$ (dash dotted magenta) and the discrete-time obtained with the interpolatory method method $\mathbf K_d$ (dashed red).}
    \label{fig:controlDisc}
\end{figure}

\begin{figure}
    \centering
    \includegraphics[width=0.49\linewidth]{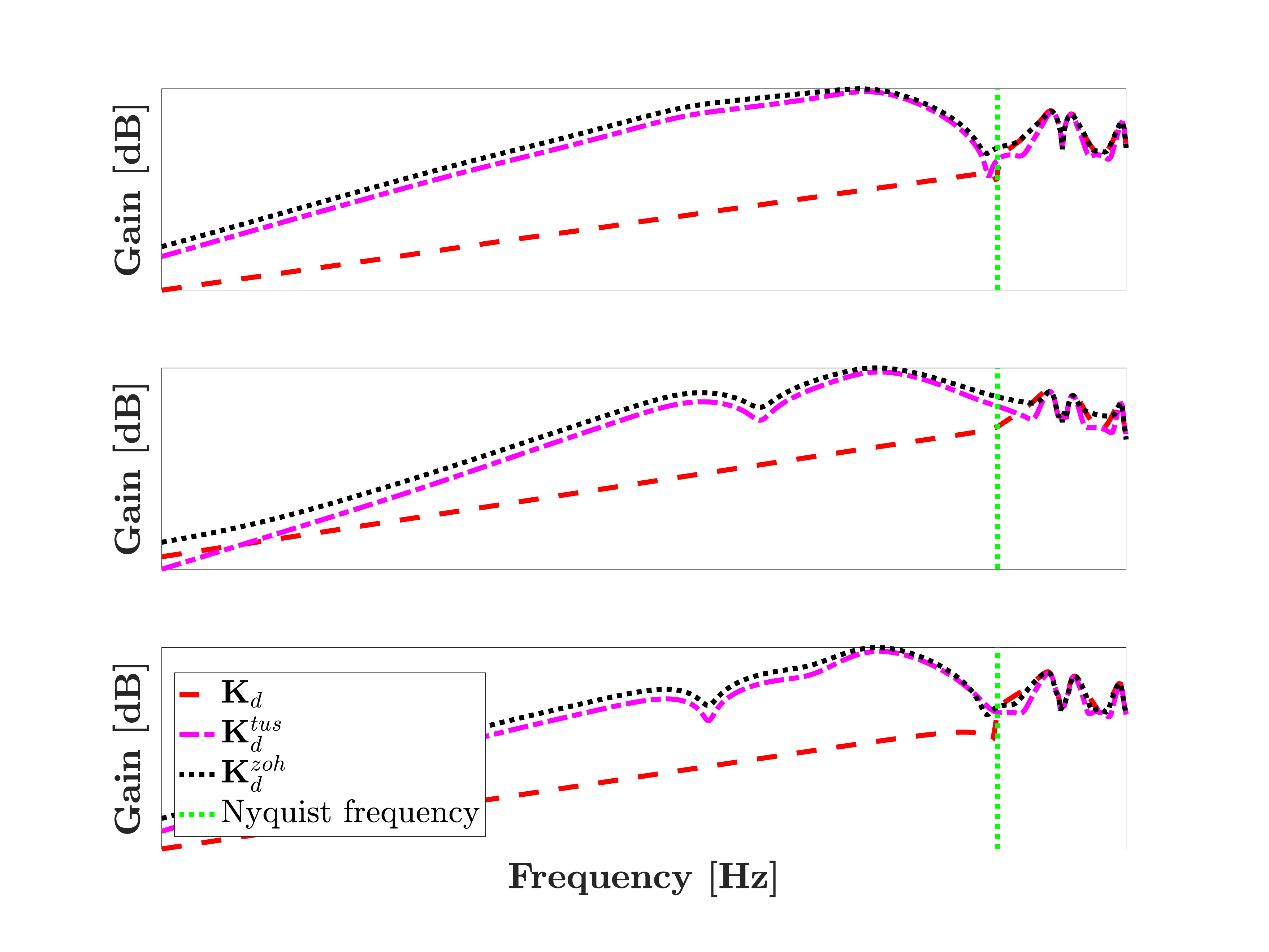}
    \includegraphics[width=0.49\linewidth]{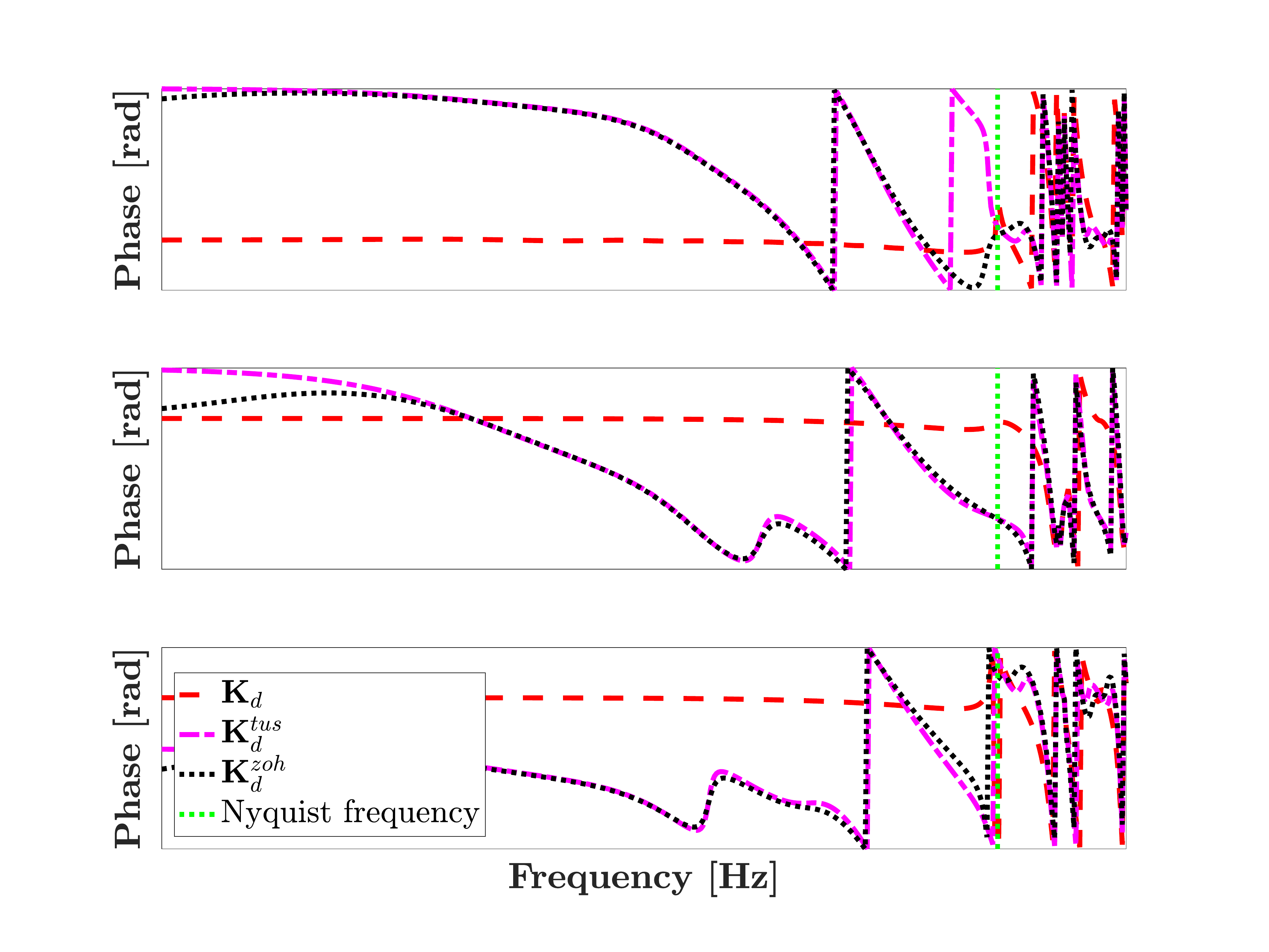}
  \caption{Gain (left) and phase (right) errors: comparison of the Bode plot of different sampled controllers with the continuous-time one $\mathbf K$  with holder. The discrete-time obtained with backward method $\mathbf K_d^{zoh}$ (dotted black), the discrete-time obtained with bilinear method $\mathbf K_d^{tus}$ (dash dotted magenta) and the discrete-time obtained with the interpolatory method method $\mathbf K_d$ (dashed red).}
    \label{fig:controlError}
\end{figure}

It should also be noted that \cref{theAlg} does not exploit the state-space structure of $\mathbf K$ and only requires frequency data from it. Therefore, as presented in the preliminary work \cite{VuilleminDiscreteArxiv}, the approach may be applied to a wider class of models than just those described by a state-space realisation. \cref{theAlg}  is applied on the obtained continuous-time GLA controller. The obtained sampled-time controller sigma plot is shown in \cref{fig:controlDisc}, comparing the responses of the continuous-time controller $\mathbf K$ interconnected with a holder, with different sampled controllers obtained with different methods (but same order $n_c$). In addition \cref{fig:controlError} reports the gain and phase mismatches of the different sampled controllers.

With reference to \cref{fig:controlDisc} and \cref{fig:controlError}, one may observe that the obtained  $\mathbf K_d$  well reproduces the original controller  $\mathbf K$ behaviour up to the Nyquist frequency, while the bilinear approach $\mathbf K_d^{tus}$  tends to compress the responses, both bilinear and backward importantly modify the phase. This accurate restitution obtained by $\mathbf K_d$ is first justified by embedding the holder model in the control law before discretisation. Second, as the complete controller is now delayed, once again the interpolatory framework is adapted and thus outperforms the classic ones. More important effects on the closed-loop will be presented in the conclusion \cref{sec:conclusions}.

%% file: figures/sd-synth.tex
%

\begin{tikzpicture}
  \node (a) at (0,0) [text width = 2.2cm,text centered] {$\mathbf H$ Continous-time plant};

  \node (pd) at ($(a.east) + (4cm,0)$) [anchor=west,text width = 2.2cm,text centered] {$\mathbf H_d$ Sampled-time plant};
  \node at ($0.5*(a.east) + 0.5*(pd.west)$) [color=gray, above] {\scriptsize Plant discretisation (zoh)};

  \node (kd) at ($(pd.south) + (0,-2cm)$) [anchor=north,text width = 2.2cm,text centered]{$\mathbf K_d$ Sampled-time controller};

  \node at ($0.5*(pd.south) + 0.5*(kd.north)$) [color=gray, right,text width=1.2cm] {\scriptsize Discrete synthesis};
  \draw [->] (a.east) -- (pd.west);
  \draw [->] (pd.south) -- (kd.north);

  \node (k) at ($(a.south) + (0,-2cm)$) [anchor=north,text width = 2.2cm,text centered]{$\mathbf K$ Continuous-time controller};

  \draw [->] (a.south) -- (k.north);
  \draw [->] (k.east) -- (kd.west);

  \node at ($0.5*(k.east) + 0.5*(kd.west)$) [color=gray,text width = 3.2cm,text centered] {\scriptsize Controller discretisation (zoh, tustin, etc.)};
  \node at ($0.5*(a.south) + 0.5*(k.north)$) [color=gray, right,text width=1.2cm] {\scriptsize Analog synthesis};

  \draw [->, dashed] ($(a.315) +(0.3cm,0)$) -- ($(kd.north)+(-0.9cm,0)$);
  \node at ($0.5*(a.315) + 0.5*(kd.135) - (0,0.2)$) [anchor = center,color=gray, above=0.2cm, rotate=-23] {\scriptsize Sampled-data synthesis};

\end{tikzpicture}


%% file: sec-conclusions.tex
In this paper, a complete end to end approach for the construction of a GLA control function has been presented, applied and implemented on a real-life generic BizJet aircraft use-case simulator. As a conclusion to this paper,  \cref{sub:perfo} presents the obtained performances when plugging the discrete-time controller in the complete industrial  simulator including the time-delayed continuous-time model (hundreds of simulations are performed). As the proposed design scheme can be readily applied to any linear dynamical systems and not only aircraft control, in \cref{sub:overview} we summarise the main methodological contributions and potential improvement axes.

\subsection{BizJet aircraft GLA function performances}
\label{sub:perfo}

\begin{figure}[htbp]
  \centering
  \includegraphics[width=.6\textwidth]{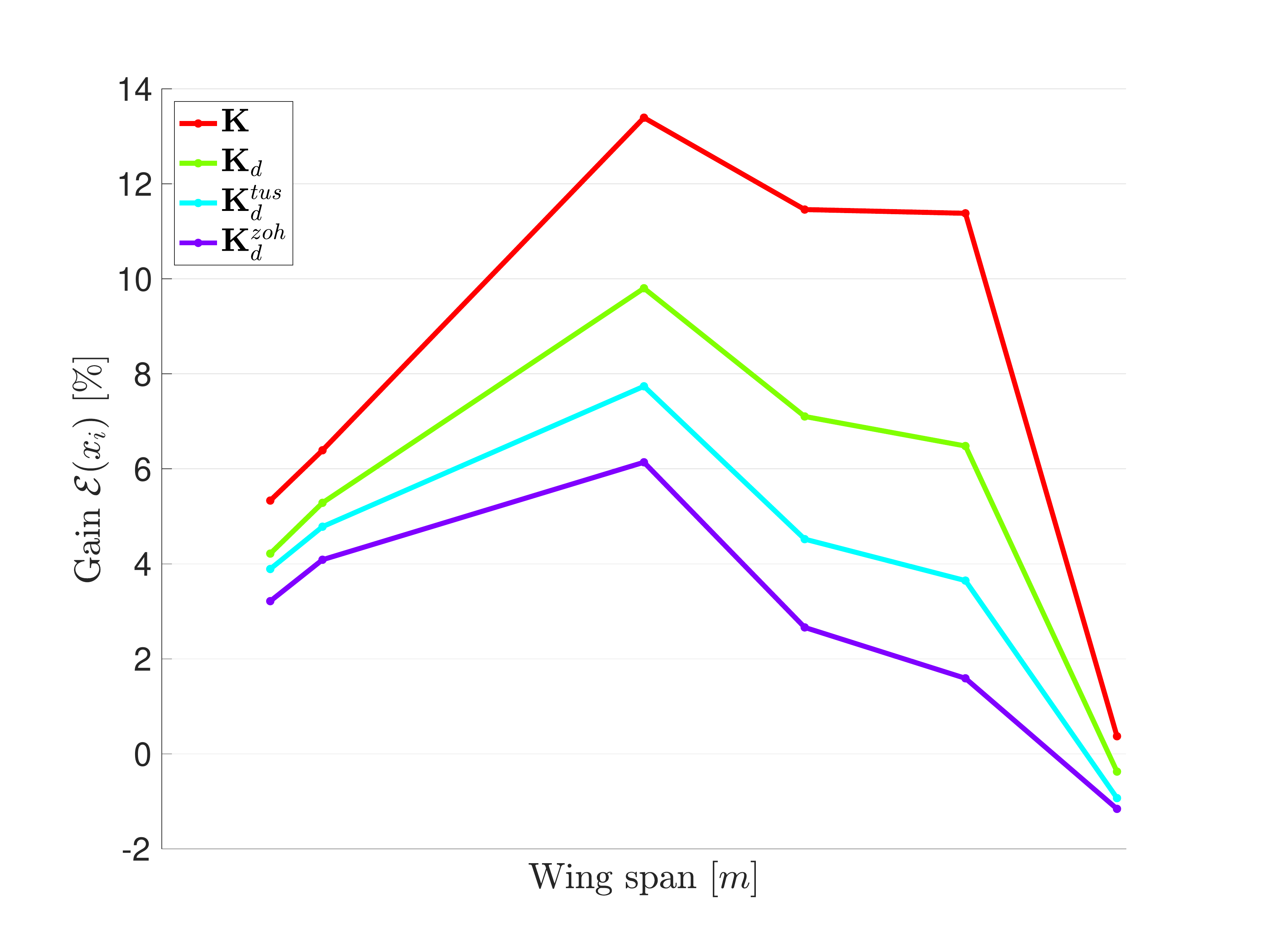}
  \caption{Gust load attenuation envelope gain $\mathcal E(x_i)$ resulting to the interconnection presented in \cref{fig:global} for different GLA controllers: the continuous-time one $\mathbf K$ (red), the discrete-time obtained with backward method $\mathbf K_d^{zoh}$ (purple), the discrete-time obtained with bilinear method $\mathbf K_d^{tus}$ (cyan) and the discrete-time obtained with the interpolatory method method $\mathbf K_d$ (green).}
\label{fig:envelope}
\end{figure}

The main objective of GLA is to reduce the so-called load envelope, \ie the load amplification along the wing span in response to a gust signal family usually considered by aircraft manufacturers and authorities. This alleviation feature is presented on \cref{fig:envelope}, for different discretised controllers, with the same dimension and sampling-time value $h$. 

\begin{figure}[htbp]
  \centering
  \includegraphics[width=.6\textwidth]{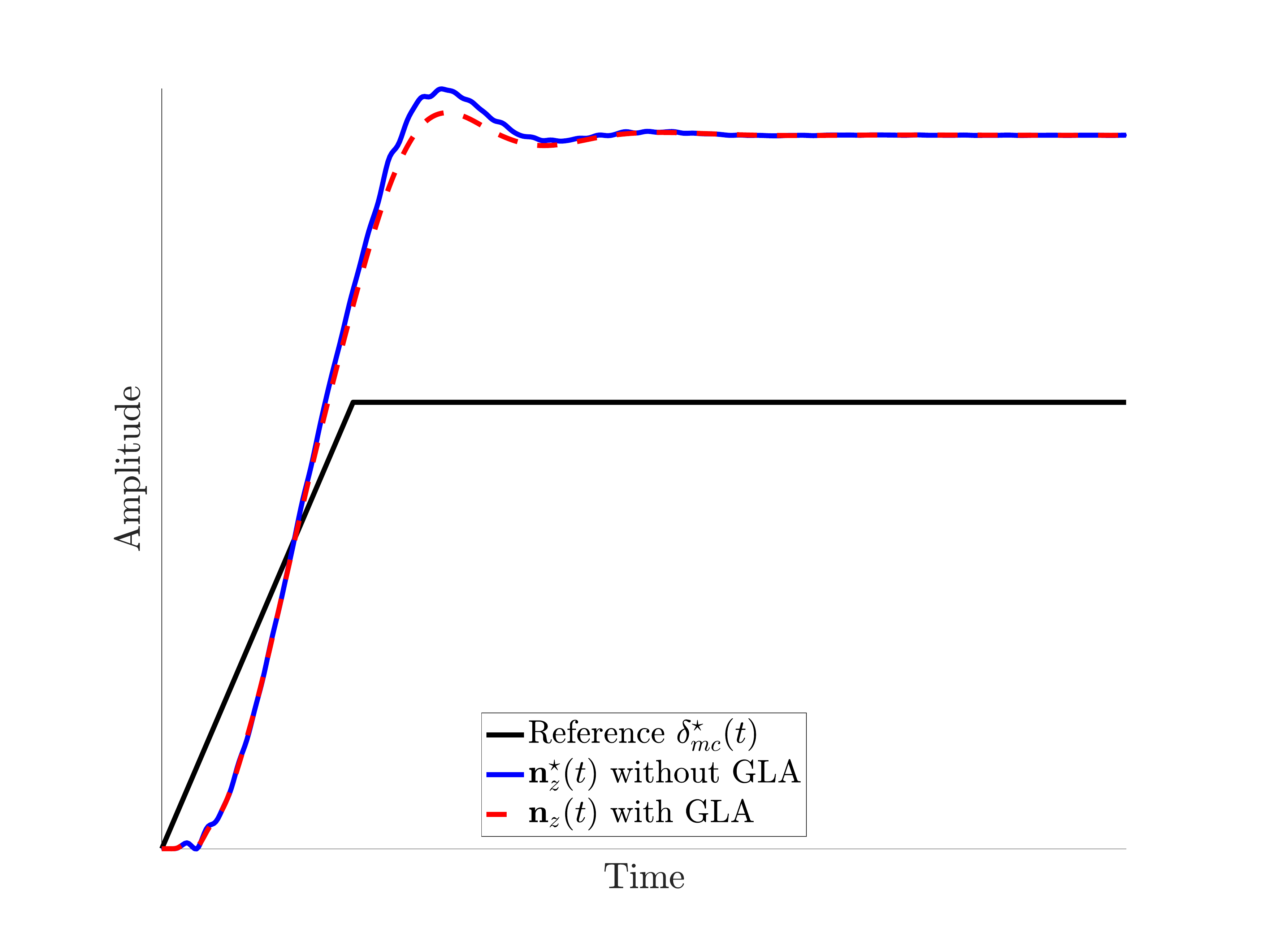}
  \caption{Time-domain response of the interconnection \cref{fig:global}  without (solid blue) and with (dashed red) the GLA function, to a pilot stick input (solid black).}
\label{fig:fq}
\end{figure}

While, the ``optimal'' gust load alleviation level obtained by the continuous-time controller $\mathbf K$ leads to attenuation between 6 to 13\% (except at the wing tip, where the loads are very small), the sampled-time versions with a sampling value $h$ fixed by the available on-board computer clearly deteriorate the performances. Still, depending on the discretisation method, one clearly observes an important gain when using the proposed discretisation  approach of \cref{theAlg}, with respect to the standard ones. It is to be noted that every additional percent may represent hundreds of kilograms of mass reduction, and thus of aircraft global consumption. In addition, as a standard industrial requirement, the proposed load alleviation should not result in a change of flight performances. This is ensured by the control design and illustrated on \cref{fig:fq} where the normalised load factor variation in response to an elevator request $\mathbf \delta_{mc} ^\star(t)$, results in similar behaviours with and without the GLA function, which fulfils the industrial process and constraints. 

In this paper, results are shown for few (worst case) configurations, but obviously, the entire process has been validated over all hundreds of cases and gust waves. This  contributes in addressing both industrial safety and quality, as well as helping in the certification process. Moreover, all the presented tools were made available in the industrial process, running robustly in the workflow.

\subsection{General contributions overview and improvements axes}
\label{sub:overview}

The paper considers the challenge of GLA control function design and implementation while handling industrial constraints. Although this application is very challenging as it involves a family of complex medium-scale delayed models and considers different control specifications, both in time and frequency domains, authors believe that the process can be  readily applied  to any kind of complex linear dynamical systems. 

Indeed, the main message of the article is to place \emph{interpolatory methods} at the center of the modelling, design and discretisation steps. The main contributions of the paper can then be viewed from three angles. First, from a \emph{methodological point of view}, the proposed process is flexible, numerically affordable and tailored to any (finite and infinite dimensional) linear dynamical models. Second, from a more \emph{theoretical point of view}, it provides a complete view of interpolatory methods for dynamical systems, applicable to both realisation-based and transfer function-based models. Finally, \cref{theAlg} gives a complete process to perform a dynamical model/controller discretisation resulting in an alternate  solution to the classical literature methods, standing then as a third, \emph{algorithmic} contribution.
